\documentclass[aps,pra,10pt,twocolumn,groupedaddress,notitlepage,showpacs,floatfix,longbibliography]{revtex4-1}
\pdfoutput=1
\usepackage{graphicx,graphics,epsfig,times,bm,bbm,amssymb,amsmath,amsfonts,amsthm,mathrsfs,MnSymbol}
\usepackage{gensymb}
\usepackage[matrix,frame,arrow]{xypic}
\usepackage[pdfstartview=FitH]{hyperref}

\usepackage[caption=false]{subfig}

\hypersetup{
    colorlinks=true,       
    linkcolor=red,          
   citecolor=magenta,        
    filecolor=magenta,      
    urlcolor=cyan,           
    runcolor=cyan
}
\usepackage[pdftex]{color}
\usepackage{braket}
\usepackage{enumerate}
\usepackage[normalem]{ulem}
\usepackage[usenames,dvipsnames]{xcolor}
\usepackage{multirow}
\usepackage{mathtools}
\usepackage{placeins}

\usepackage{nameref}
\usepackage{varioref}
\usepackage{hyperref}
\usepackage{cleveref}

\definecolor{orange}{rgb}{1,0.5,0}

\newcommand{\bes} {\begin{subequations}}
\newcommand{\ees} {\end{subequations}}
\newcommand{\bea} {\begin{eqnarray}}
\newcommand{\eea} {\end{eqnarray}}

\definecolor{gold}{rgb}{0.85,.66,0}

\newcommand{\beq}{\begin{equation}}

\newcommand{\eeq}{\end{equation}}

\newcommand{\ignore}[1]{}

\def\tr{\mathrm{Tr}}

\def\s{\sigma}

\def\>{\rangle}
\def\<{\langle}

\newcommand{\Mod}[1]{\ (\mathrm{mod}\ #1)}

\def\s0{I}

\newcommand{\ig}[1]{}

\begin{document}
\title{Coined quantum walk on a quantum network}
\author{Jigyen Bhavsar$^*$, Shashank Shekhar$^*$, Siddhartha Santra\\\textit{$^1$ Department of Physics and Center of excellence in Quantum information, computation, science and technology, Indian Institute of Technology Bombay, Mumbai, Maharashtra 400076, India}\\ $^*$ contributed equally}
\begin{abstract} 
We explore a discrete-time, coined quantum walk on a quantum network where the coherent superposition of walker-moves originates from the unitary interaction of the walker-coin with the qubit degrees of freedom in the quantum network. The walk dynamics leads to a growth of entanglement between the walker and the network on one hand, and on the other, between the network-qubits among themselves. The initial entanglement among the network qubits plays a crucial role in determining the asymptotic values of these entanglement measures and the quantum walk statistics. Specifically, the entanglement entropy of the walker-network state and the negativity of the quantum network-qubit state saturate to values increasing with the initial network-entanglement. The asymptotic time-averaged walker-position probability distribution shows increasing localization around the initial walker-position with higher initial network entanglement. A potential application of these results as a characterisation tool for quantum network properties is suggested.

\end{abstract}
\maketitle

\section{Introduction}

Quantum walks are a model of quantum dynamics where a walker can follow a superposition of paths on a graph instead of randomly and incoherently picking among the different possible paths at various stages of the evolution as occurs in a classical random walk \cite{spitzer2001principles,Rudnick_Gaspari_2004,norris1998markov}. Due to their coherent exploration of possible paths and faster spreading \cite{konno2002quantum,kendon2003decoherence} such walks can be utilised to design powerful algorithms \cite{childs2009universal,lovett2010universal,santha2008quantum}, generate entanglement for various applications \cite{carneiro2005entanglement,berry2011two,rohde2012entanglement,maloyer2007decoherence}, simulate physical processes \cite{mohseni2008environment}, modeling the quantum-to-classical transition \cite{brun2003quantum} and have been experimentally realised in a variety of physical systems ranging from single-photons in space \cite{broome2010discrete} to linear ion-traps \cite{schmitz2009quantum}. 

In canonical models of  discrete-time coined quantum walks (DCQW) \cite{ambainis2001one,nayak2000quantum}, three separate systems are involved: the quantum walker, the walker's coin and the underlying graph on which the walk takes place. At every time-step, a fixed unitary coin-operator acts on the coin resulting in a coin-state which is a superposition over its basis states representing the possible moves of the walker. A subsequent unitary operator, also fixed, shifts the walker to positions conditioned on the coin-state resulting in corresponding superpositions of the various possible moves. The underlying graph in this case plays a passive role and, essentially, provides a set of positions for the walker to move. 

A quantum network \cite{q_net_kimble}, on the other hand, is a collection of interconnected, possibly entangled, quantum systems - typically designed to enable a variety of useful distributed quantum information processing tasks \cite{Wehnereaam9288}. Mathematically, a quantum network can be described as a graph, $G(V,E)$, whose vertices, $V$, house the separated quantum systems while the graph edges, $E$, represent bipartite entanglement between some of those systems \cite{mondal2024entanglement}. Visualising a quantum network in this manner raises the intriguing possibility of using them as the underlying graph for a novel type of quantum walk - a quantum walk on a quantum network. The quantum network in this case can play a fundamentally different, \emph{active}, role in the walk-evolution than in canonical quantum walks, since, in addition to supplying a set of vertices as walker positions, it provides network quantum degrees of freedom that can coherently interact with the quantum-walker leading to richer quantum-walk behavior.

In this work, we explore a coined quantum walk on a quantum network whose evolving quantum state, due to the walk dynamics, generates the coherent superposition of walker-moves via controlled-unitary interactions with the quantum-walker at each discrete time-step. At the start of the walk, the quantum-walker is in a product state with the quantum network qubits which are initially entangled among themselves in a specific manner. Then, the overall unitary walk dynamics produces entangled walker-network states at subsequent times. The reduced dynamics of the quantum-walker is, therefore, not unitary - contrasting with the situation in canonical quantum walks \cite{andraca_review, kadian_review}.

We study the quantum-walk statistics, in terms of the walker-position probability distribution, via its asymptotic time-averaged stationary distribution and find that it reflects the magnitude of the initial entanglement in the quantum network. Interestingly, a higher magnitude of the initial network entanglement corresponds to a higher localization of the stationary distribution around the initial position of the walker on the network graph. The rate of approach to this asymptotic distribution, however, is independent of the initial network entanglement, depending only on the network size, although higher initial network-entanglement values lead to more fluctuations in the approach. Further, we verify that the walk dynamics indeed leads to interference between certain amplitudes and therefore fulfils the criteria for quantum walks \cite{kendon_qw_criteria}. On the quantum network side, the walk dynamics creates time-evolving quantum correlations between previously uncorrelated network qubits. As a measure of this correlation at long-times, we study the asymptotic bipartite entanglement among the network qubits via the negativity \cite{vidal_negativity} of the network-qubit state, which also saturates to values that increase with the initial network entanglement. Finally, based on our observations for this novel model of a quantum walk on a quantum network, that, its time-dependent and asymptotic statistics bear signatures of the initial network entanglement we point out a potential application of such walks to probe certain entanglement properties of quantum networks, such as the average entanglement over its edges. 

The rest of the paper comprises of three main sections. In Sec. \ref{background} we provide the necessary background for canonical discrete-time coined quantum walks in Subsec. \ref{subsec:DCQW-L} and quantum networks in Subsec. \ref{subsec:qnets}. Then, we describe a coined quantum walk on a quantum network in Sec. \ref{sec:QWQN-L} including a description of the initial network entanglement in Subsec. \ref{subsec:INE}, time-dependent quantum walk state in Subsec. \ref{subsec:time_dep_state}, time-dependent walker-position probability distribution in Subsec. \ref{subsec:time_dep_walker_position}, the quantum walk on a quantum network as a superposition of conditional quantum walks in Subsec. \ref{subsec:cond_q_walk}, the long-time average probability distribution in Subsec. \ref{subsec:stationary_dist}, approach to the stationary distribution in Subsec. \ref{subsec:approach_asymptotic}, entanglement evolution in Subsec. \ref{subsec:entropy}, interference among the walk state amplitudes in Subsec. \ref{subsec:interference} and walk statistics as a probe of network entanglement in Subsec. \ref{subsec:network_probe}. Finally, we conclude with a discussion and scope for future work in Sec. \ref{sec:disc_conc}.

\section{Background}
\label{background}
\subsection{Discrete-time coined quantum walk on a line (DCQW-L)}
\label{subsec:DCQW-L}
In a DCQW-L the quantum walk takes place on an underlying linear graph, $G(V,E)$, with the set of graph vertices and edges denoted $V$ and $E$ respectively. Canonically, two quantum degrees of freedom, the walker and the coin, participate in a DCQW-L. The coin is a quantum system internal to the walker whose Hilbert space, $\mathcal{H}_c$, is spanned by two states, $\{\ket{0}_c, \ket{1}_c\}$, in case of the linear graph $G$. The walker's Hilbert space, $\mathcal{H}_p$, is spanned by the set of position eigenstates, \{$\ket{n}_{p}$\}, where, $n\in V=\{1,2,...,N\}$, is the set of positive integers that label the vertices of $G$. The time-dependent joint-state of the walker and the coin is an element of the tensor product Hilbert-space, $\mathcal{H}=\mathcal{H}_c\otimes \mathcal{H}_p$.

The discrete time-evolution of walker-coin state can be described as iterations of a two-stage process. In the first stage, a one-qubit unitary coin-operator, for example, the Hadamard gate $\hat{H}_c$, is applied to the coin. In the second stage, a conditional unitary shift-operator, $\hat{S}$, with the coin-state as the control and walker-position as the target is applied on the joint coin-walker state. The overall unitary, $\hat{U}$,  of this two-stage process for a single time-step is,
\begin{align}
\hat{U}=\hat{S}_{c\to p}(\hat{H}_c\otimes \openone_p),
\label{one-step-u}
\end{align}
where, the operators $\hat{H}_c$ and $\hat{S}$ are given by,
\begin{align}
\hat{H}_c&=\frac{1}{\sqrt{2}}(\ket{0}\bra{0}_{c} + \ket{0}\bra{1}_{c} + \ket{1}\bra{0}_{c} - \ket{1}\bra{1}_{c})\nonumber\\
\hat{S}_{c\to p}&=\ket{0}\bra{0}_{c}\otimes\sum_{n}\ket{n+1}\bra{n}_{p} + \ket{1}\bra{1}_{c}\otimes\sum_{n}\ket{n-1}\bra{n}_{p}.
\label{operators-DCQW-L}
\end{align}

The joint-state of the coin and the walker, $\ket{\phi(t)}$, after $t\in \mathbb{Z}^+$-time steps starting from an initial state, $\ket{\phi}_I$, is given by,
\begin{align}
\ket{\phi(t)} = (\hat{U})^{t}\ket{\phi}_{I}
\label{t-step-state}
\end{align}
with $\hat{U}$ given by Eq.~(\ref{one-step-u}). Typically, the initial state of the coin and walker is assumed to be an unentangled state, given by $\ket{\phi}_{I} =\ket{c_I}_c \otimes  \ket{n_I}_{p}$ which is understood as starting the walk with the walker localized at the vertex $n_I\in V$ of the graph with the coin state initialized to $\ket{c_I}$.

The unitary evolution of the walker-coin state as indicated by Eq.~(\ref{t-step-state}) results in a time-dependent state of the form,
\begin{align}
\ket{\phi(t)} = \sum_{n}[a_{n}(t)\ket{0}_{c} + b_{n}(t)\ket{1}_{c}]\otimes \ket{n}_{p},
\label{state-qw}
\end{align}
where, $a_n(t),b_n(t)\in \mathbb{C}$, the sum over $n$ is over all the vertex labels of $G$ and the unitarity of the dynamics implies $\sum_n |a_n(t)|^2+|b_n(t)|^2=1\forall t$. The walk dynamics therefore leads to a spreading of the probability amplitude  over the different vertices of the graph $G$ with time.

\begin{figure}
    \centering
    \includegraphics[height=6cm,width=\columnwidth]{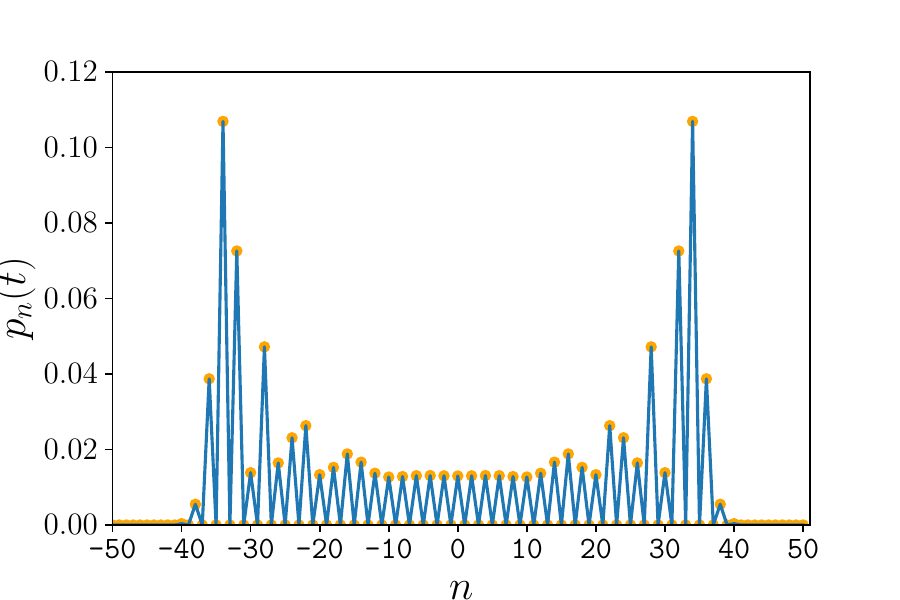}
    \caption{Walker-position probability distribution, $p_n(t)$,of a quantum walker to be found at different positions, $n$, for a discrete-time coined quantum walk on an infinite-line with a Hadamard coin operator after, $t=50$, time-steps. The coin and walker start with initial condition: $\ket{\phi}_I=\frac{1}{\sqrt{2}}(\ket{0}_{c} + i\ket{1}_{c})\otimes\ket{0}_{p}$.}
    \label{fig:Hwalk_line}
\end{figure}

Of particular interest, under DCQW-L dynamics, is the probability, $p_n(t)$, of finding the walker at a particular graph vertex, $n\in V$, at a given time, $t$. This can be obtained as the expectation value of the operator, $\Pi_n:=\ket{n}\bra{n}_p$, with respect to the state in Eq.~(\ref{t-step-state}) and is given by,
\begin{align}
p_n(t)=\bra{\phi(t)}\Pi_n\ket{\phi(t)}.
\end{align}
Obtaining, $p_n(t)$, is possible analytically when there is sufficient symmetry in the system, for example, in quantum walks on graphs with vertices of fixed degree. The translational invariance of the dynamics, over the vertices of the graph, then leads to conserved momentum which can be exploited to write the transition matrix for the amplitudes of the state Eq.~(\ref{t-step-state}) in the Fourier transformed basis in the Schrodinger approach \cite{andraca_review}. Alternately, in the combinatorial, path-integral approach, the amplitude of a position state can be obtained by tracking and adding, separately for the $\ket{0}_c$ and $\ket{1}_c$-states, the contributions that reach the position, $n$, starting from the initial position, $n_I$, via paths of length, $t$. 

Statistically, the mean position of the quantum walker at any time, for the unbiased walk described by Eq. (\ref{operators-DCQW-L}) and (\ref{t-step-state}), remains unchanged, $\braket{n(t)}=\sum_n p_n(t)n=0\forall t$. However, the variance of its position, $\braket{n^2(t)}=\sum_n p_n(t)n^2\sim t^2$, for an infinite underlying linear graph. This is distinct from the behavior of the walker in classical random walks where, $\braket{n^2(t)}\sim t$. In effect, a classical random walk is diffusive since in that case, $\sigma_{n(t)}\sim \sqrt{t}$ , whereas quantum random walks are ballistic, $\sigma_{n(t)}\sim t$, with more spreading of the walker as a function of time. One can observe this spreading behavior in DCQW-L as the time-dependent, symmetric, double-peaked probability distribution of the walker-position with the peaks close to the extremities, $n\sim \pm t$, of the probability distribution over the linear graph, Fig.~\ref{fig:Hwalk_line}.

For a DCQW-L with periodic boundary conditions, the probability distribution of the walker-position, $p_n(t)$, shows distinct behavior dependent on the time-scale, $t$ , relative to the graph size, $|V|$. For times smaller than the graph size,  $t\leq|V|$, the probability distribution is similar to the DCQW-L on an infinite line as shown in Fig.~\ref{fig:Hwalk_line}. For times larger than the size of the graph,  $t\geq|V|$, the probability distribution is typically quasi-periodic  which implies that the unitary walk operator in the R.H.S. of Eq. (\ref{t-step-state}) can become arbitrary close to the identity for certain time steps, $t$, that is, for $\epsilon\geq 0$, there is a timestep $t$ such that $|| \hat{U}^{t} - I || \leq \epsilon$. The time-dependent state, $\ket{\phi(t)}$, at such times can be arbitrarily close to the initial state of the walker, $\ket{\phi}_I$. 

Similar to the DCQW-L, in the novel model of a quantum walk on a quantum network studied in this paper, Sec. \ref{sec:QWQN-L}, the time-dependent walk state, Eq. (\ref{eq:QWQN-state-t}), is unitarily related to the initial-state in that situation and, in principle, can show quasi-periodic behavior. However, since the unitary evolution in the latter case is over the whole system, that is walker plus quantum network, the reduced dynamics of the walker degrees of freedom does not show quasi-periodic behavior. Instead, we find that the time-averaged walker-position probability distribution, $p_n(t)$, asymptotically approaches a stationary probability distribution, $\pi_n(\alpha)$, which is determined by the value of the initial network entanglement as explained in Subsec. \ref{subsec:entropy}.


\subsection{Quantum Networks}
\label{subsec:qnets}

A quantum network can be described mathematically using an underlying graph, $G(V,E)$, where the quantum systems reside at the set of vertices, $V$, with the set of edges, $E=\{(v_1,v_2),(v_1,v_5),...\}$, representing physical connections, for example photonic channels, between the vertices, $v_i\in V$. Networks can have different physical topologies, such as random or scale-free topologies or regular ones such as linear, square, hexagonal etc., depending on the degree distribution of the underlying graph, $G(V,E)$. The central characteristic of a quantum network relevant for quantum walks, distinct from the underlying graph in the standard quantum walk scenario described above, is that the qubits at different vertices in the quantum network may be entangled. Here, we will consider networks with initial entanglement over single edges, $e\in E$, of the network, $G(V,E)$. The quantum walk dynamics on such a quantum network correlates even those qubits that are not direct neighbors resulting in increasing entanglement between subsets of the quantum network qubits as we discuss in Subsec. \ref{subsec:entropy}.

\begin{figure}
    \centering
    \includegraphics[width=1\linewidth]{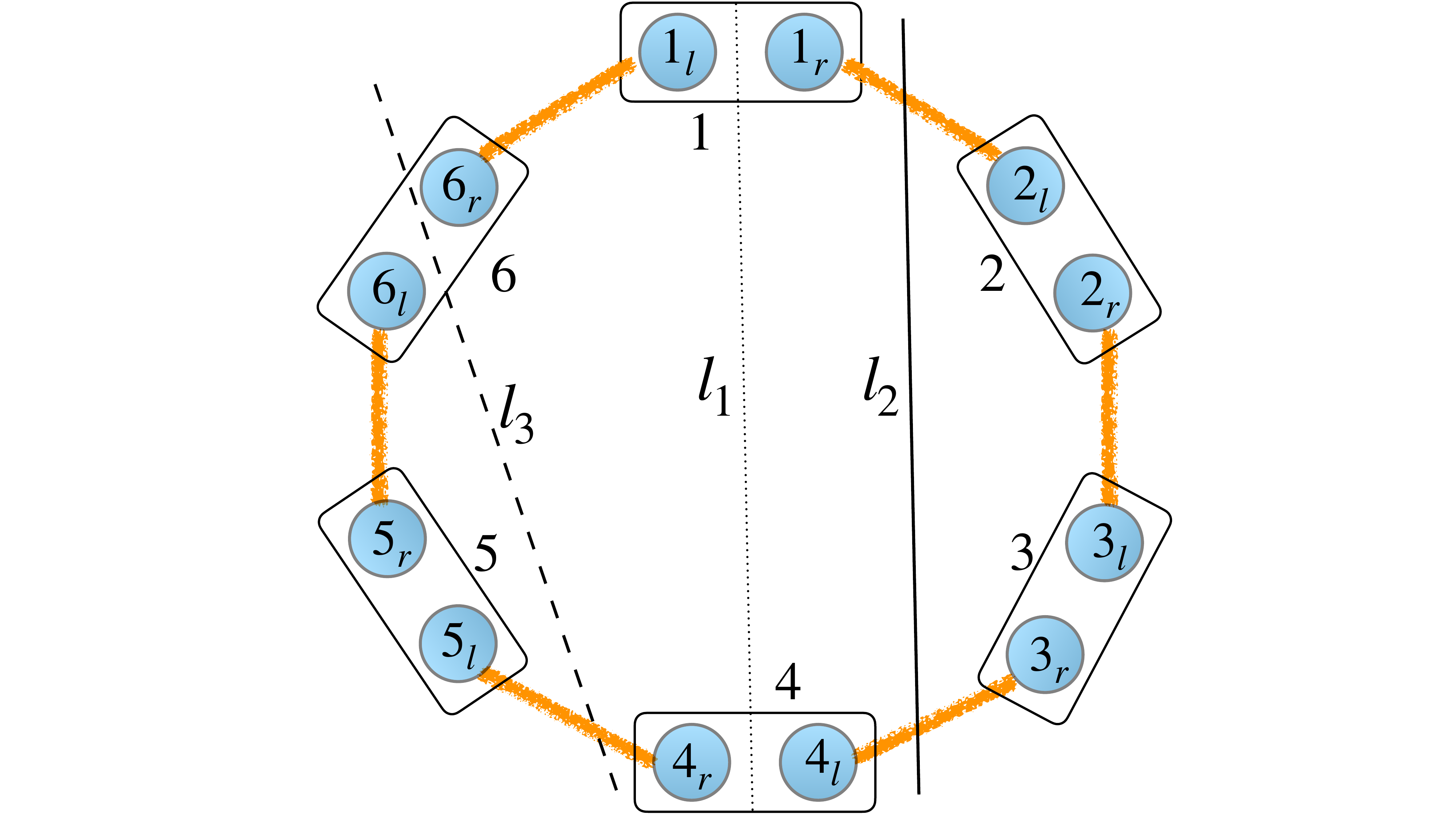}
    \caption{(Color online) A quantum network with 6 nodes and 12 qubits prior to commencement of the quantum walk. Every vertex, $i=\{1,2,...,6\}$, has two qubits (Blue) with labels, $l,r$, that are initially unentangled with each other. The two qubits at any node are in an entangled pure state (Orange thick line), $\ket{\alpha}$, with another qubit at an adjacent node. The lines $l_1,l_2,l_3$ show possible bipartitions of the system. $l_1$ denotes a bipartition where there is no entanglement between the qubits on the two sides. $l_2$ denotes a biparition which intersects two copies of the state $\ket{\alpha}$. $l_3$ denotes a bipartition which intersects one copy of the state $\ket{\alpha}$.}
    \label{fig:QN}
\end{figure}

In this work, we specialise to a quantum network on a finite linear graph, $G_L(V,E)$, with periodic boundary conditions as shown in Fig. \ref{fig:QN}. In this network, every vertex, $v_i\in V$, in the set of vertices, $V$, is connected to the nearest neighbor vertices on either side. The number of edges is, therefore, $|E|=|V|$. Further, every vertex houses two qubits - each of which are in two separate entangled pure states with another qubit at the neighboring node on either side.  We assume that initially the network comprises of a total of $|E|$-entangled pure states, $\ket{\psi}_e, e\in E$, of every pair of qubits sharing an edge in $G_L(V,E)$ with the initial state of the quantum network permitting the form,
\begin{align}
\ket{\psi}_G=\otimes_{e\in E}\ket{\psi}_e.
\end{align}

In such a quantum network with pure states of the form, $\ket{\psi}_e=\ket{\alpha_e}=\sqrt{\alpha_e}\ket{00}+\sqrt{1-\alpha_e}\ket{11},~\forall e\in E$, along its edges we define and utilise two notions of entanglement among the qubits of the network. The first is that of the initial network entanglement, $INE(\ket{\psi}_G)$ defined via,
\begin{align}
INE(\ket{\psi_G}):=\frac{\sum_{e\in E}\alpha_e}{|E|},
\end{align}
that is, as the average of the squares of the Schmidt coefficients of the two-qubit entangled pure states over the network edges. The quantity in the right hand side of the above equation stands as a proxy for the network entanglement because for the state on any edge, the Von Neumann entanglement entropy \cite{review_area_laws} of the state, $S_{VN}(\ket{\alpha})=-\alpha\log(\alpha)-(1-\alpha)\log(1-\alpha)$, increases monotonically with $\alpha$ in the interval $\alpha\in[0,0.5]$. We limit our consideration of the states $\ket{\alpha}$ to the range $\alpha\in[0,0.5]$, since the quantum walk dynamics remains invariant under a global spin flip ($\ket{0}\leftrightarrow \ket{1}$ of each network qubit), resulting in the transformation of the state, $\ket{\alpha}\to\ket{1-\alpha}$, over any edge.

We use the term, homogenous, when referring to quantum networks with identical states, $\ket{\alpha_e}=\ket{\alpha}\forall e\in E$, along its edges. Conversely, the term, inhomogenous, when referring to networks with states that are different for the various edges. For homogenous networks, we can write the initial state of the quantum network as,
\begin{align}
\ket{\psi}_G=\otimes_{e\in E}\ket{\psi}_e=\ket{\alpha}^{\otimes|E|},
\end{align}
in which case, the initial network entanglement is given by,
\begin{align}
INE(\ket{\alpha}^{\otimes |E})=\alpha.
\end{align}
For inhomogenous networks, the initial network entanglement, $\overline{\alpha}=INE(\otimes_{e\in E}\ket{\psi}_e)$, is an average of the $\alpha_e$ values over the different edges. We remark that most of our analytical work focuses on homogenous quantum networks whereas inhomogenous networks are treated numerically here.

The second notion of entanglement in the quantum network is that of bipartite network entanglement, $BNE_{A|B}(\rho_G)$, when considering two disjoint subsets of the network qubits, $A,B$, with $A\cup B=V, A\cap B=\emptyset$ for a general state of the quantum network qubits, $\rho_{G}$. We define the bipartite network entanglement as the negativity \cite{vidal_negativity} of state, $\rho_G$,
\begin{align}
BNE_{A|B}(\rho_G):=\mathcal{N}_{A|B}(\rho_G)=\frac{||(\rho_G)^{T_A}||_1-1}{2},
\end{align}
where, $||(\rho_G)^{T_A}||_1$ is the trace-norm of the partial transpose of $\rho_G$ with respect to the subsystem $A$. Clearly, the value of $BNE_{A|B}(\rho_G)$ depends on the choice of the bipartition, for example, a bipartition provided by a line crossing two vertices and dividing the two qubits at each vertex into opposite subsets ($l_1$ in Fig. \ref{fig:QN}) will result in a zero value of the $BNE_{A|B}(\rho_G)$ at the initial time. Whereas the bipartitions provided by the lines $l_2,l_3$ in the same figure have an initial non-zero value of the $BNE_{A|B}(\rho_G)$. We will see later that the initial network entanglement, $INE(\ket{\psi_G})$, determines the statistics of the quantum walker degrees of freedom as well as the saturation value of the asymptotic ($t\to\infty$) bipartite network entanglement, $BNE_{A|B}(\rho_G)$. Numerically, we study the growth of the negativity between the qubits of the network across a bipartition of the $l_1$-type in Subsec. \ref{subsec:entropy}. 

\section{Coined quantum walk on a quantum network}
\label{sec:QWQN-L}
We now consider a discrete-time coined quantum walk on a linear quantum network (QWQN-L) with periodic boundary conditions. The quantum network on which the quantum walk takes place has identical pure entangled states $\ket{\alpha}$ over each of its edges, that is, it a homogenous network in the state, $\ket{\psi}_G=\ket{\alpha}^{\otimes |E}$. For a network of, $|V|=N$, vertices with two qubits at each vertex, the total Hilbert space dimension of the quantum network, walker-coin and the walker position, $\mathcal{H}_{QWQN}$, is: $\text{Dim}(\mathcal{H}_{QWQN})=\text{Dim}(\mathcal{H}_G\otimes \mathcal{H}_c\otimes \mathcal{H}_p)=2^{2N}\times 2\times N=N2^{2N+1}$.

The initial coin state is assumed to be, $\ket{c_0}_c=(\ket{0}_c+i\ket{1}_c)/\sqrt{2}$, and the walker is assumed to start from the initial position, $n=0$, which corresponds to the initial position state, $\ket{0}_p$. Thus, the overall factorized state at the initial time is,
\begin{align}
\ket{\psi(0)}=\ket{\psi}_G\otimes \ket{c_0}_c\otimes \ket{0}_p,
\label{eq:qwqn-state-0}
\end{align}
where all three degrees of freedom, viz. the quantum network qubits, the walker-coin and the walker-position, are unentangled; further, the network qubits are also only pairwise entangled among themselves along the network edges, $E$.

Each time-step of the QWQN-L comprises of two stages analogous to those in the DCQW-L. In the first stage, the quantum walker's coin degree of freedom undergoes a local, coherent interaction, via a conditional Hadamard, $\hat{H}_{G\to c}$, with the qubits at the nodes of the network - resulting in a transformation of the coin state conditioned on the state of the network qubits. Explicitly, the operator $\hat{H}_{G\to c}$ can be expressed as,  
\begin{align}
\hat{H}_{G\to c}&= \sum_{n\in V} \hat{H}_{n_l\to c}\hat{H}_{n_r\to c}\nonumber\\
&= \sum_{n\in V} [(\ket{0}\bra{0}_{n_l}\otimes \openone_c+\ket{1}\bra{1}_{n_l}\otimes \hat{H}_c)\nonumber\\
&~~~~~~~~~~~~~~~~~~~~~~~(\ket{0}\bra{0}_{n_r}\otimes \openone_c+\ket{1}\bra{1}_{n_r}\otimes \hat{H}_c)]\nonumber\\
&=\sum_{n\in V}(\ket{00}\bra{00}_{n_ln_r}+\ket{11}\bra{11}_{n_ln_r})\otimes \openone_c\nonumber\\
&~~+\sum_{n\in V}(\ket{01}\bra{01}_{n_ln_r}+\ket{10}\bra{10}_{n_ln_r})\otimes \hat{H}_c\nonumber\\
&=\sum_{n\in V}(P^e_{n_ln_r}\otimes \openone_c+P^o_{n_ln_r}\otimes \hat{H}_c)
\label{coinop}
\end{align}
where, the operators 
\begin{align}
P^e_{n_ln_r}&=\ket{00}\bra{00}_{n_ln_r}+\ket{11}\bra{11}_{n_ln_r},\nonumber\\
P^0_{n_ln_r}&=\ket{01}\bra{01}_{n_ln_r}+\ket{10}\bra{10}_{n_ln_r},
\end{align}
measure the parity of the qubits at the node, $n\in V$. The $\hat{H}_{G\to c}$ operator therefore applies a Hadamard-operator to the walker-coin in case the qubits at a node are measured to have odd-parity but leaves it unchanged otherwise.
Note, also that the transformation, $\hat{H}_{G\to c}$, entangles the coin degree of freedom with the network qubits and thus after the first time-step the quantum network state is no longer the pure state, $\ket{\psi}_G$.

\begin{figure}
    \centering
    \includegraphics[height=7.5cm,width=.85\columnwidth]{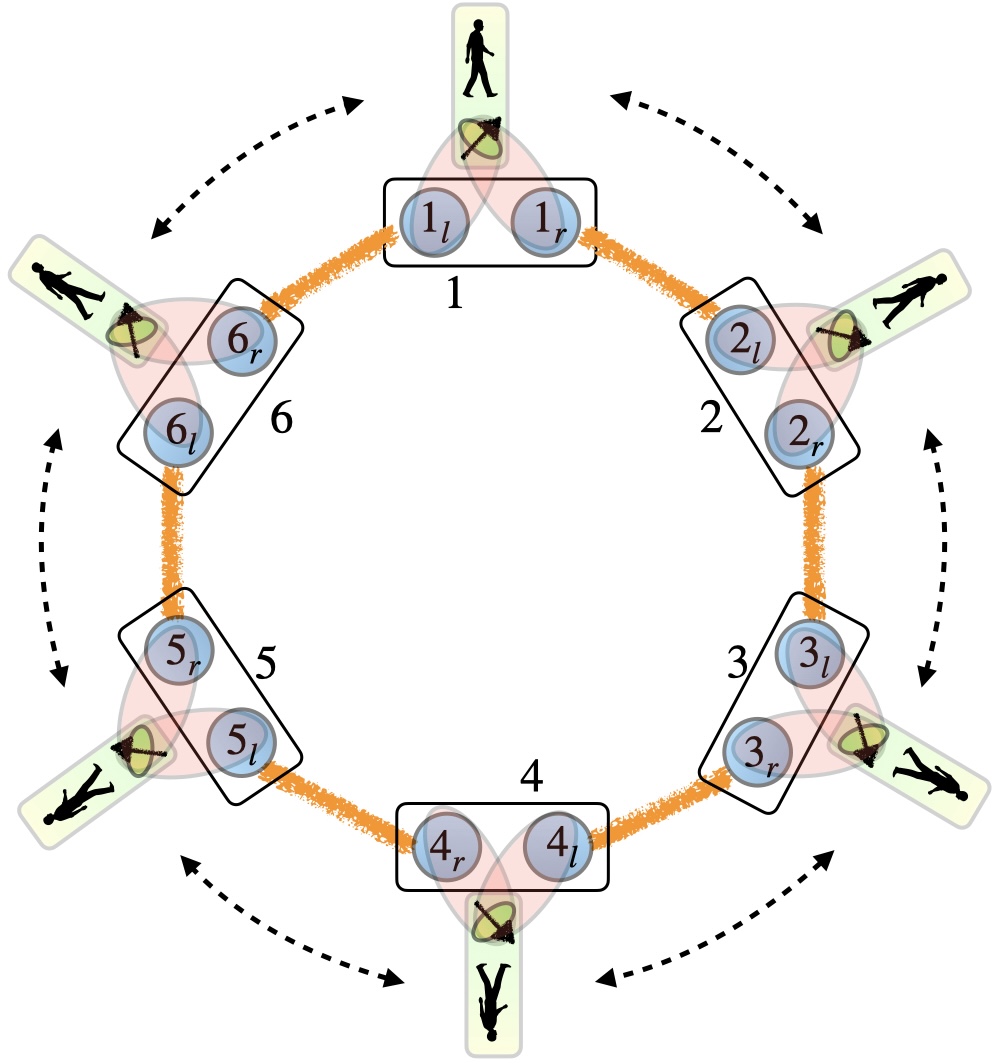}
    \caption{Visualization of a coined quantum walk on a linear quantum network with periodic boundary conditions. At the initial time, at each vertex there are two unentangled qubits which are in a pure entangled state with another qubit on either side. At every time-step, two controlled-Hadamard gates are applied on the coin state with the vertex qubits as control followed by a shift-operator conditioned on the coin-state.}
    \label{fig:1DQN}
\end{figure}

In the second stage, the usual shift operator, $\hat{S}_{c\to p}$, from Eq. (\ref{operators-DCQW-L}) as in the DCQW-L case, moves the walker to a coherent superposition of position states depending on the superposition state of the coin qubit while leaving the network qubits untouched. The total unitary operator corresponding to a single time-step of the QWQN-L is therefore,
\begin{align}
\hat{U}_{QWQN}= (\openone_G\otimes \hat{S}_{c\to p} )(\hat{H}_{G\to c}\otimes \openone_p),
\label{eq:u-qwqn}
\end{align}
with the walker state after $t$ - time-steps given by,
\begin{align}
\ket{\psi(t)}=(\hat{U}_{QWQN})^t\ket{\psi(0)}
\label{eq:QWQN-state-t}
\end{align}
which is a state that generally correlates all three degrees of freedom. Moreover, it creates correlations between previously uncorrelated quantum network qubits among themselves as we shall see.

\subsection{Initial network entanglement and walker-move bias}
\label{subsec:INE}
The bias of the quantum walker-moves at any node, $n\in V$, depends on the concurrence, $C(\ket{\alpha})$, as can be seen by considering the first time-step behavior of the QWQN-L through the action of the operator, $\hat{U}_{QN}$. The two qubits, $0_l,0_r$, at the vertex, $n=0$, participate in two separate but identical entangled states, $\ket{\alpha}=\sqrt{\alpha}\ket{00}+\sqrt{1-\alpha}\ket{11}$. Thus, the parity measurements have the following outcome probabilities,
\begin{align}
p^e_{0_l0_r}&= \bra{\alpha}^{\otimes 2} P^e_{0_l0_r}\otimes \openone_4 \ket{\alpha}^{\otimes 2}=\alpha^2+(1-\alpha)^2\nonumber\\
p^o_{0_l0_r}&= \bra{\alpha}^{\otimes 2} P^o_{0_l0_r}\otimes \openone_4 \ket{\alpha}^{\otimes 2}=2\alpha(1-\alpha),
\label{eq:prob_0}
\end{align}
with, $p^e_{0_l0_r}+p^o_{0_l0_r}=1$, and the $\openone_4$ acts on the two qubits of the state $\ket{\alpha}^{\otimes 2}$ other than $0_l,0_r$. The state of the coin conditioned on the outcomes of the parity measurements of the network qubits, $0_l,0_r$, at the starting position is, therefore,
\begin{align}
&\text{Even parity of}~(0_l,0_r):\ket{c_0}\to\hat{\openone}\ket{c_0},\nonumber\\
&\text{Odd parity of}~(0_l,0_r):\ket{c_0}\to\hat{H}\ket{c_0},
\label{eq:parity_operators}
\end{align}
which implies that a computational basis state of the coin, $\ket{c_0}=\{\ket{0}_c,\ket{1}_c\}$, stays unchanged when the $(0_l,0_r)$-qubits have even parity but transforms to an equal superposition when the parity is odd. The subsequent action of the shift operator, $\hat{S}_{c\to p}$, in the second stage, therefore, shifts the walker position in the same direction as the previous time-step for the even parity situation but splits into a coherent superposition of left and right walker moves for the odd parity situation.

Importantly, notice that the probability of the odd outcome, $p^o_{0_l0_r}$, equals the Concurrence, $C(\ket{\alpha})=2\alpha(1-\alpha)$, of the state, $\ket{\alpha}$, which is an entanglement measure for pure states \cite{entformation}. The concurrence, $C(\ket{\alpha})$, monotonically increases with $\alpha\in[0,0.5]$ and monotonically decreases for $\alpha\in[0.5,1]$. Therefore, for the quantum network described above with maximally entangled two-qubit pure states ($\alpha=0.5$) on its edges, $p^o_{0_l0_r}=p^e_{0_l0_r}=0.5$, implying equal probability of left- and right- walker-moves at the initial position. On the other hand, for a network with completely unentangled qubits ($\alpha=0,1$) we get that, $p^o_{0_l0_r}=0,~p^e_{0_l0_r}=1$ in which case the walker moves in the direction determined by the initial coin state, $\ket{c_0}$.

The direct relationship of the entanglement of the state, $C(\ket{\alpha})$, with the bias of the walker-moves at the first time-step becomes less apparent for subsequent time-steps. This is because for further time-steps, the entanglement of the quantum network qubits among themselves does not permit a simple calculation of the probabilities, $p^e_{n_ln_r},p^o_{n_ln_r}$ as in Eq. (\ref{eq:prob_0}) for an arbitrary vertex, $n\in V$. Instead, for subsequent time-steps, $t\geq 2$, the probabilities of the even and odd parity measurements at a vertex, $n\in V$, are obtained more generally by,
\begin{align}
p^{e,o}_n(t)&= \tr[ P^{e,o}_{n_ln_r} \rho_{n_ln_r}(t)],
\end{align}
where, the expectation value is taken with respect to the time-dependent state, 
\begin{align}
\rho_{n_ln_r}(t)=\tr_{\{V\backslash \{n_l,n_r\}\}\cup c\cup p}[\ket{\psi(t)}\bra{\psi(t)}],
\end{align}
of the qubits, $n_l,n_r$, obtained as their reduced density matrix when the global state of the quantum walk is, $\ket{\psi(t)}\in \mathcal{H}_{QWQN}$. The notation, $\tr_{\{V\backslash \{n_l,n_r\}\}\cup c\cup p}[.]$, denotes the partial trace with respect to all qubits of the network other than, $n_l,n_r$, and the walker-coin and walker position. The probability of the parity outcomes, and thus of the walker moves, is therefore dependent on the position, $n$, the entanglement parameter, $\alpha$, and the time, $t$.

\subsection{Time-dependent state of the quantum walk on the quantum network}
\label{subsec:time_dep_state}

The time-dependent state, $\ket{\psi(t)}\in \mathcal{H}_{QWQN}=\mathcal{H}_G\otimes \mathcal{H}_c\otimes\mathcal{H}_p$, jointly of the quantum network-qubits, walker-coin and walker-position, as given by Eq. (\ref{eq:QWQN-state-t}), starting from the product initial state given in Eq. (\ref{eq:qwqn-state-0}) is a superposition over products of the quantum network states $\ket{G_i}$, the walker-coin states $\{\ket{0}_c,\ket{1}_c\}$, and the walker-position states, $\ket{n}_p$. The state of the quantum walk at any time is thus given by,
\begin{align}
\ket{\psi(t)}&=(\hat{U}_{QWQN})^t [\ket{\alpha}^{\otimes N}\otimes \frac{\ket{0}_c+i\ket{1}_c}{\sqrt{2}}\otimes \ket{0}_p]\nonumber\\
&=\sum_{n}[ (\sum_i a^i_{n}(t)f_i(\alpha) \ket{G_i})\ket{0}_{c} \nonumber\\
&~~~~~~~~~~~~~~~~~~~~~~~~~~~~~~~+ (\sum_i b^i_{n}(t) f_i(\alpha) \ket{G_i})\ket{1}_{c}]\otimes \ket{n}_{p}\nonumber\\
\label{state-qwqn-t}
\end{align}
where, the sum over, $n$, is over all the walker position states, $n=1,2,...,|V|=N$, and the index, $i$, is summed over, $i=0,1,2,...,2^{|V|}-1$, which label the computational basis states of the quantum network qubits, $\ket{G_i}$. Note that the total number of computational basis states of the $2N$ network-qubits is actually $2^{2N}$. However, every adjacent pair of network-qubits is initially in the superposition state, $\ket{\alpha}=\sqrt{\alpha}\ket{00}+\sqrt{1-\alpha}\ket{11}$ and thus have even parity which is conserved under the action of the unitary operator, $\hat{U}_{QWQN}$, given in Eq. (\ref{eq:u-qwqn}). The effective Hilbert-space dimension of every pair of entangled qubits is therefore $2$ - which reduces the Hilbert-space dimension of the $2N$ network-qubits to $2^{N}$. Thus, only these many computational basis states of the quantum network contribution to the superposition in Eq. (\ref{state-qwqn-t}).

The relevant computational basis states, $\ket{G_i}$, for, $i=0,1,2,...,(2^{N}-1)$, of the network qubits can be understood as the state corresponding to the $2N$-bit binary representation of the number, $i$, when each bit is duplicated next to itself. As an example in a network with, $N=3$, we have, 
\begin{align}
\ket{G_0}&=\ket{000000}, \ket{G_1}=\ket{000011}, \ket{G_2}=\ket{001100},\nonumber\\
\ket{G_{3}}&=\ket{001111}, \ket{G_{4}}=\ket{110000}, \ket{G_{5}}=\ket{110011}.\nonumber\\
\ket{G_{6}}&=\ket{111100}, \ket{G_{7}}=\ket{111111}.
\label{comp_basis_N3}
\end{align}
Thus, for general, $N$, one can use the representation,
\begin{align}
\ket{G_i}=\otimes_{j=0}^{(N-1)}\ket{x_{2j}}\ket{x_{2j+1}},
\label{comp_basis}
\end{align}
where, $x_{2j}=x_{2j+1}$ is the $j$-th bit in the $N$-bit binary representation of the number, $i$. Clearly, these states are orthonormal, that is, $\braket{G_i,G_j}=\delta_{i,j}$. 

Let us now notice, from equation (\ref{state-qwqn-t}) above, that in the time-dependent state of the QWQN the coefficients of the position and coin states are the (unnormalised) superpositions of computational basis states, $\ket{G_i}$, of the network qubits which we call the network states, $\ket{\mathcal{G}^{0,1}_n(\alpha,t)}$, given by,
\begin{align}
\ket{\mathcal{G}^0_n(\alpha,t)}&=\sum_{i=0}^{(2^N-1)}a^i_n(t)f_i(\alpha)\ket{G_i}\nonumber\\
\ket{\mathcal{G}^1_n(\alpha,t)}&=\sum_{i=0}^{(2^N-1)}b^i_n(t)f_i(\alpha)\ket{G_i},
\label{state-graphsuper}
\end{align}
with,
\begin{align}
\sum_{i=0}^{(2^N-1)} f_i^2(\alpha)&=1\nonumber\\
\sum_{n=1}^{N} |a^i_n(t)|^2+|b^i_n(t)|^2&=1,\forall i, t.
\label{f_a_b}
\end{align}

Thus, we can write the state, $\ket{\psi(t)}$, using Eqs. (\ref{state-qwqn-t}) and (\ref{state-graphsuper}) as,
\begin{align}
\ket{\psi(t)}=\sum_{n=1}^{N}[ \ket{\mathcal{G}^0_n(\alpha,t)}\ket{0}_{c}+ \ket{\mathcal{G}^1_n(\alpha,t)}\ket{1}_{c}]\otimes \ket{n}_{p},
\label{state-graphsuper-2}
\end{align}
which is written in an analogous manner to the walk state, (\ref{state-qw}),  in the standard DCQW-L. 

The time-dependent state, Eq. (\ref{state-graphsuper-2}), in the QWQN-L has a fundamentally different behavior from that for the DCQW-L where the coefficients, $a_n(t), b_n(t)$, were simple scalar complex numbers. In the QWQN-L, the walker dynamics creates time-dependent correlations between configurations of the network qubits represented by the network states, $\ket{\mathcal{G}^{0}_n(\alpha,t)}$ and $\ket{\mathcal{G}^{1}_n(\alpha,t)}$, which appear as the coefficents of the product of walker-coin and walker-position states, $\ket{0}_c\otimes \ket{n}_p$ and  $\ket{1}_c\otimes \ket{n}_p$. 

Let us now note that the network states, $\ket{\mathcal{G}^{0}_n(\alpha,t)}$ and $\ket{\mathcal{G}^{1}_n(\alpha,t)}$, determine the action of the conditional Hadamard operator, $\hat{H}_{G\to c}$, on the walker-coin. Since different walker-position states, $\ket{n}_p$, are in a tensor product with distinct network states in Eq. (\ref{state-graphsuper-2}), we find that the action of $\hat{H}_{G\to c}$ acting on the walker-coin differs based on the walker-position. The QWQN-L, therefore, lacks the translational invariance of walk-operator action over the network vertices unlike the situation in the DCQW-L. This implies that Fourier transform based methods \cite{andraca_review} to determine the time-dependent functions $a_n^i(t),b_n^i(t)$ are not useful in the case of the QWQN-L, see Appendix (\ref{app:ft}).

\subsection{Time-dependent walker-position probability in the QWQN-L}
\label{subsec:time_dep_walker_position}
The probability of finding the walker at a specific position, $n\in V$, on the  quantum network after time, $t$, is thus,
\begin{align}
p_n(t)&=\bra{\psi(t)}(\ket{n}\bra{n}_p)\ket{\psi(t)}\nonumber\\
&=\braket{\mathcal{G}^0_n(\alpha,t)|\mathcal{G}^0_n(\alpha,t)}+\braket{\mathcal{G}^1_n(\alpha,t)|\mathcal{G}^1_n(\alpha,t)}\nonumber\\
&=\sum_{i=0}^{(2^N-1)} (|a^i_n(t)|^2+|b^i_n(t)|^2)f_i^2(\alpha),
\label{prob_n_t}
\end{align}
with the quantities $a_n^i(t),b_n^i(t),f_i(\alpha)$ obeying the normalization conditions given by Eq. (\ref{f_a_b}). Note that, while $p_n(t)\leq 1\forall n\in V$, we have, $\sum_{n}p_n(t)=1\forall t$, using Eq.~(\ref{f_a_b}) in Eq. (\ref{prob_n_t}) above. Further, note that at time, $t=0$, we have, $p_{n}(t=0)=\delta_{n,m}$, if the walker is initialised at the, $m\in V$, position.

\begin{figure}[htp]
\subfloat[Low initial network entanglement, $\alpha=0.1$.]{%
  \includegraphics[clip,width=\columnwidth]{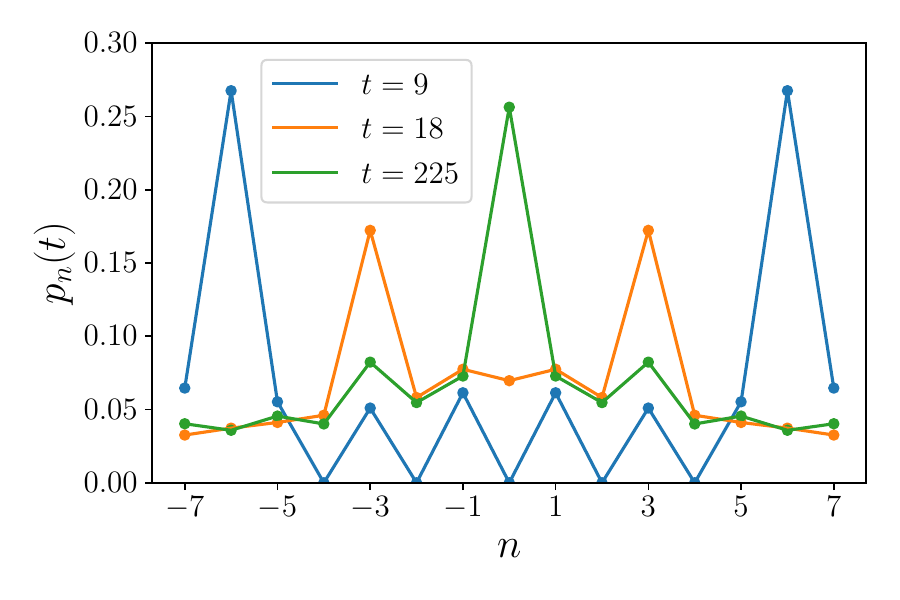}%
}

\subfloat[Maximum initial network entanglement, $\alpha=0.5$.]{%
  \includegraphics[clip,width=\columnwidth]{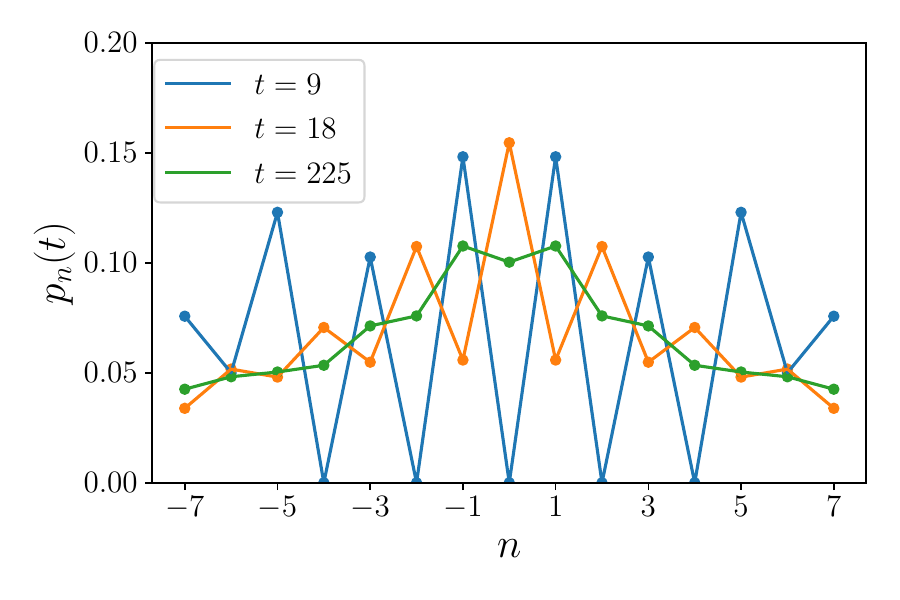}%
}
\caption{Walker-position probability distribution, $p_{n}(t)$, with initial coin-state, $(\ket{0}_{c} + i\ket{1}_{c})/\sqrt{2}$, initial walker-position state, $\ket{0}_p$ in a network with $|V|=15$ vertices. The top panel (a) and bottom panel (b) respectively show the distribution for low and high initial network entanglement at times smaller ($t=9$), comparable ($t=20$) and much larger ($t=225$) than the system size, $V$.}
\label{prob_t_N15}
\end{figure}

It is instructive to observe the behavior of $p_n(t)$ numerically as shown in Fig. \ref{prob_t_N15} for a concrete realisation of a QWQN-L. The figure shows the behavior of $p_n(t)$ from Eq. (\ref{prob_n_t}) for the initial coin state, $(\ket{0}_c+i\ket{1}_c)/\sqrt{2}$, and initial walker-position, $n=0$, on a linear quantum network with $|V|=15$ vertices with, $n\in\{-7,-6,...,0,...,6,7\}$ and periodic boundary conditions. The subfigures reveal the dynamical behaviour of the walker at different time scales and for different initial network entanglement scenarios. The top and bottom subfigures, respectively, depict the behavior for low and high initial network entanglement, $\alpha=0.1$ and $\alpha=0.5$, at times smaller ($t=9$), comparable ($t=18$) and much larger ($t=225$) than the quantum network size, $|V|$. 

The top panel in Fig. \ref{prob_t_N15} shows that for low initial entanglement, $\alpha=0.1$, the time-dependent probability distribution, $p_n(t)$, has two primary peaks at the positions, $n \equiv r \Mod{n}$, which move ballistically in opposite directions starting from the initial vertex, $n=0$. The other vertices in the network have rather low probabilities though the secondary probability peaks closer to the initial vertex are higher. This implies that upon time-averaging the stationary probability distribution shows some localization around the initial position as shown in Fig. \ref{fig:stat_dist_1}.

The bottom panel in Fig. \ref{prob_t_N15} shows that for high entanglement, $\alpha=0.5$, the walker-position probability distribution, $p_n(t)$, stays localized around the initial vertex for long times, $t\sim \mathcal{O}(|V|)$ and $t\gg \mathcal{O}(|V|)$, though for times smaller than the system size, $t<|V|$, the distribution is more uniform. In fact, the localization around the initial vertex increases with the initial network entanglement, $\alpha$, as is reflected in the time-averaged stationary probability distribution shown in Fig. \ref{fig:stat_dist_1}.

\subsection{Quantum walk on an quantum network as a superposition of conditional quantum walks}
\label{subsec:cond_q_walk}
The QWQN-L state, $\ket{\psi(t)}$, admits an interesting interpretation as a coherent superposition of states corresponding to independent quantum walks - each conditioned on a specific computational basis state of the quantum network qubits.
Rewriting the state, $\ket{\psi(t)}$, from Eq. (\ref{state-qwqn-t}), by shifting the time-dependence away from the network states we get, 
\begin{align}
\ket{\psi(t)}&=\sum_{i} f_{i}(\alpha) \ket{G_i}\otimes [\sum_n (a^i_n(t)\ket{0}_c+b^i_n(t)\ket{1}_c)\otimes \ket{n}_p],\nonumber\\
&=\sum_{i} f_{i}(\alpha) \ket{G_i}\otimes\ket{W_i(t)},
\label{physical_pic_state}
\end{align}
with, $\sum_{i}f_i^2(\alpha)=1$, showing that the state of the QWQN-L is a normalised superposition of independent quantum walk states, $\ket{W_i(t)}\in\mathcal{H}_c\otimes \mathcal{H}_p$, with,
\begin{align}
\ket{W_i(t)}&=\sum_n (a^i_n(t)\ket{0}_c+b^i_n(t)\ket{1}_c)\otimes \ket{n}_p,\nonumber\\
\text{and},&\sum_{n}|a_n^i(t)|^2+|b_n^i(t)|^2=1 \forall i, t,
\label{walk_states}
\end{align}
where, each quantum walk state, $\ket{W_i(t)}$, is conditioned on, $\ket{G_i}\in\mathcal{H}_G$, which is a computational basis state of the network qubits, described via Eq. (\ref{comp_basis_N3}) and (\ref{comp_basis}), and which remain invariant under the action of the quantum walk operator, $\hat{U}_{QWQN}$, given in Eq. (\ref{eq:u-qwqn}). Further, these conditional walk states are normalised but not orthogonal,
\begin{align}
\braket{W_i(t)|W_i(t)}&=\sum_n |a^i_n(t)|^2+|b^i_n(t)|^2=1,\forall i, t,\nonumber\\
\braket{W_i(t)|W_j(t)}&=\sum_n a^i_n(t)a^j_n(t)+b^i_n(t)b^j_n(t)\neq 0.
\label{walk_states_og}
\end{align}
Finally, the functions, $f_i(\alpha)$, can be determined from Eq. (\ref{state-qwqn-t}) to be the terms in the binomial series, $(\sqrt{\alpha}+\sqrt{1-\alpha})^N$, with,
\begin{align}
f_i(\alpha)=\alpha^{\text{wt}(i)/2}(1-\alpha)^{(N-\text{wt}(i))/2},
\label{f_i}
\end{align}
where, $\text{wt}(i)$ is the Hamming weight of the $N$-bit binary representation of, $i\in \{0,1,2,...,(2^N-1)\}$.

The physical picture that emerges from Eqs. (\ref{physical_pic_state})-(\ref{f_i}) for the quantum walk on a quantum network, therefore, is one where the share of the total probability, $f_i^2(\alpha)$, for any of the conditional quantum walks, $\ket{W_i(t)}$, is determined by the configuration of network qubits in the computational basis state, $\ket{G_i}$. The amplitudes of the position states, $\ket{n}_p$, interfere within each of the independent walk states, $\ket{W_i(t)}$, but not across different walks, $\ket{W_i(t)},\ket{W_j(t)},i\neq j$, since they are coupled to orthogonal computational basis states, $\ket{G_i},\ket{G_j}$, of the network qubits. The conditional quantum walks

The alternate form of the QWQN-L state, $\ket{\psi(t)}$, written in Eq. (\ref{physical_pic_state}), reveals that the time-dependent state of the quantum network qubits, $\rho_G(t)=\tr_{c,p}[\ket{\psi(t)}\bra{\psi(t)}]$, can be represented by the density matrix,
\begin{align}
\rho_G(t)&=\tr_{c,p}[\ket{\psi(t)}\bra{\psi(t)}]\nonumber\\
&=\sum_{i,j=0}^{(2^N-1)}f_i(\alpha)f_j(\alpha)\braket{W_j(t)|W_i(t)}\ket{G_i}\bra{G_j}.
\label{time-dep-network_state}
\end{align}
The diagonal elements, $(\rho_G)_{ii}(t)=f_i^2(\alpha)$, are thus invariant with time since, $\braket{W_i(t)|W_i(t)}=1$, from Eq. (\ref{walk_states_og}), whereas the off-diagonal elements, $(\rho_G)_{ij}(t)=f_i(\alpha)f_j(\alpha)\braket{W_j(t)|W_i(t)}$, do vary with time. Further, from the general relation between density matrix elements, $|(\rho_G)_{ij}(t)|^2\leq (\rho_G)_{ii}(t)(\rho_G)_{jj}(t))$, we get that, $|\braket{W_j(t)|W_i(t)}|^2\leq 1\forall t$.

The reduced density matrix, $\rho_W(t)$, of the quantum walker and the coin can then also be obtained from Eq. (\ref{physical_pic_state}) as,
\begin{align}
\rho_W(t)&=\tr_G[\ket{\psi(t)}\bra{\psi(t)}]\nonumber\\
&=\sum_{i=0}^{(2^N-1)}f_i^2(\alpha)\ket{W_i(t)}\bra{W_i(t)}
\label{eq:coin-position_denmat}
\end{align}


\subsection{Long-time average probability distribution}
\label{subsec:stationary_dist}
The probability distribution, $p_n(t)$, of the walker-position does not converge to a limiting distribution \cite{aharonov2002quantum} at large times, $t\to \infty$, due to the overall unitary dynamics of the QWQN-L. However, the time-averaged walker-position probability distribution, $\overline{{p}_{n}(T)}$, approaches a stationary distribution, $\pi_{n}(\alpha)$, with,
\begin{align}
\pi_{n}(\alpha) =\lim_{T\to \infty} \frac{1}{T}\sum_{t=1}^T p_{n}(t),
\end{align}
that is determined by the initial network entanglement, $INE(\alpha)$, or equivalently, by the value of $\alpha$.

\begin{figure}[h]
    \centering
    \includegraphics[height=5.75cm,width=\columnwidth]{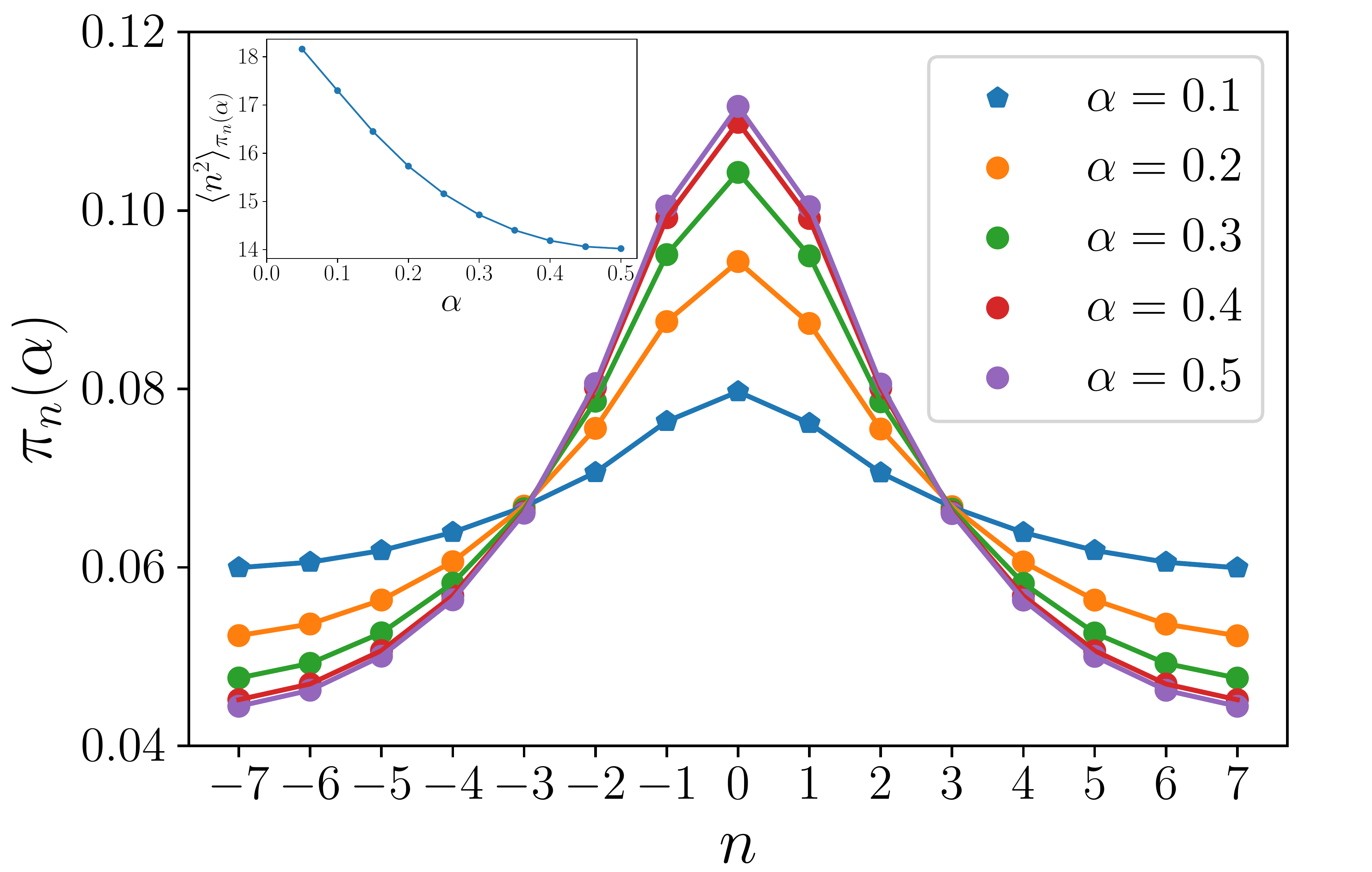}
    \caption{Stationary probability distribution, $\pi_n(\alpha)$, for the walker-position, $n$, in the QWQN-L for different values of the initial network entanglement, $\alpha$, in a homogenous network. The quantum network has $N=15$ nodes with periodic boundary conditions, and an unbiased initial coin state, $\ket{c_0}_c=(\ket{0}_{c} + i\ket{1}_{c})/\sqrt{2}$ and starts from the initial position, $n=0$. The inset shows the variance of the walker position, $\sigma_n^2=\braket{n^2}_{\pi_n(\alpha)}-\braket{n}^2_{\pi_n(\alpha)}$.}
    \label{fig:stat_dist_1}
\end{figure}
The stationary distribution, $\pi_{n}(\alpha)$, can be interpreted as the fraction of time the walker spends at a given vertex, $n\in V$, when the walker-position is sampled over a long period of time. 

We use two alternative methods to obtain the stationary distribution, $\pi_n(\alpha)$. In the first method, suitable for smaller systems $N\sim 10$, we numerically diagonalise the full unitary operator for the quantum walk on the network, $\hat{U}_{QWQN}$ shown in Eq. (\ref{eq:u-qwqn}), for a given size of the network $N$ and its initial entanglement, $\alpha$. Since, $\hat{U}_{QWQN}$ acts on the full state, $\ket{\psi(t)}$ of the network, walker-position and walker-coin, it is a $(N2^{2N+1}\times N2^{2N+1})$-dimensional matrix. Assuming its eigenvalues and eigenvectors to be, $\{e^{i\theta_l},\ket{U_l}\}_l,l=1,2,...,N2^{2N+1}$, we can expand the initial state of the system, $\ket{\psi_I}=\sum{a_k} \ket{U_k}$ in the eigenbasis of $\hat{U}_{QWQN}$ and write the time-dependent state in the same basis as, $\ket{\psi(t)}=\sum_{l,k}a_ke^{i\theta_l t}\ket{U_l}\braket{U_l|U_k}=\sum_{l}a_le^{i\theta_lt}\ket{U_l}$.
The probability, $p_n(t)$, is then be obtained as,
\begin{align}
p_n(t)=\sum_{k,l}e^{-i(\theta_k-\theta_l)t} a_ka^*_l\bra{U_l}(\openone_{G}\otimes\openone_{c}\otimes \ket{n}\bra{n}_p)\ket{U_k},
\end{align}
with the long-time average given by terms with, $\theta_k=\theta_l$, that is,
\begin{align}
\pi_n(\alpha)=\lim_{T\to\infty}\overline{p_n(t)}=\sum_{l} |a_l|^2\bra{U_l}(\openone_{G}\otimes\openone_{c}\otimes \ket{n}\bra{n}_p)\ket{U_l},
\label{eq:stat_dist_whole}
\end{align}
which assumes non-degeneracy of the phases of the eigenvalues $\theta_l$ - a fact borne out by numerical results as shown in Fig. \ref{fig:distance} which shows that the time-averaged probability distribution of the walker-position reaches an asymptotic value with negligible fluctuations.

In the second method for determining, $\pi_n(\alpha)$, we use the physical insight given by Eq. (\ref{physical_pic_state}) for the state of the QWQN-L as a superposition of conditional quantum walks, $\ket{W_i(t)}$. As previously described in Subsec. \ref{subsec:cond_q_walk}, the walk states $\ket{W_i(t)}$ appear in tensor products with corresponding orthogonal network basis state, $\ket{G_i}$, in the state, $\ket{\psi(t)}=\sum_{i} f_{i}(\alpha) \ket{G_i}\otimes\ket{W_i(t)}$. The amplitudes of the walker-coin and walker-position states, $\ket{c}\otimes\ket{n}_p$, in each, $\ket{W_i(t)}$, therefore do not interfere with the corresponding amplitudes for any other $\ket{W_j(t)}$ where $j\neq i$. Thus, the walker-position probability is the weighted sum of the probabilities at a given position, $n$, for all of these independent walks with $f^2_{i}(\alpha)$ as the weights. This is also true for the stationary distribution,
\begin{align}
\pi_n(\alpha)=\sum_{i=0}^{(2^N-1)}f_i^2(\alpha)\pi_n^i(\alpha).
\end{align}

This gives us a recipe for efficiently calculating the stationary distribution even for large networks where diagonalizing $\hat{U}_{QWQN}$ might be computationally challenging. Instead, let us notice that the states, $\ket{G_i(t)}$, are invariant under the QWQN-L unitary operator $\hat{U}_{QWQN}$ - so its action on any of the states in the supersposition can be considered as a conditional walk unitary operator, $\hat{U}_i$, such that,
\begin{align}
\hat{U}_{QWQN}\ket{G_i}\otimes\ket{W_i(t)}=(\openone_G\otimes\hat{U}_i)\ket{G_i}\otimes\ket{W_i(t)},
\label{eq:u_i}
\end{align}
with $\hat{U}_i$ acting only on the walk state, $\ket{W_i(t)}$. The dimension of each such $\hat{U}_i$ is $(2N\times 2N)$ which are thus efficiently diagonalizable even for large, $N\gg 10$. The long-time walker-position probability distribution for each independent conditional walk, $\pi^{i}_n(\alpha)$, can then be calculated in analogy with Eq. (\ref{eq:stat_dist_whole}) as,
\begin{align}
\pi^i_n(\alpha)=\sum_{j}|a^i_j|^2\braket{U^i_j|(\openone_c\otimes \ket{n}\bra{n}_p)|U^i_j},
\end{align}
where, $\ket{U^i_j}$ are the eigenvectors of $\hat{U}_i$ and the $a^i_j$ are the expansion coefficients of the initial joint-state of the walker-coin and walker-position in these bases, $\ket{c_0}\otimes\ket{n_0}=\sum_{j} a^i_j \ket{U^i_j}$.

\begin{figure}[h]
    \centering
    \includegraphics[height=6cm,width=\columnwidth]{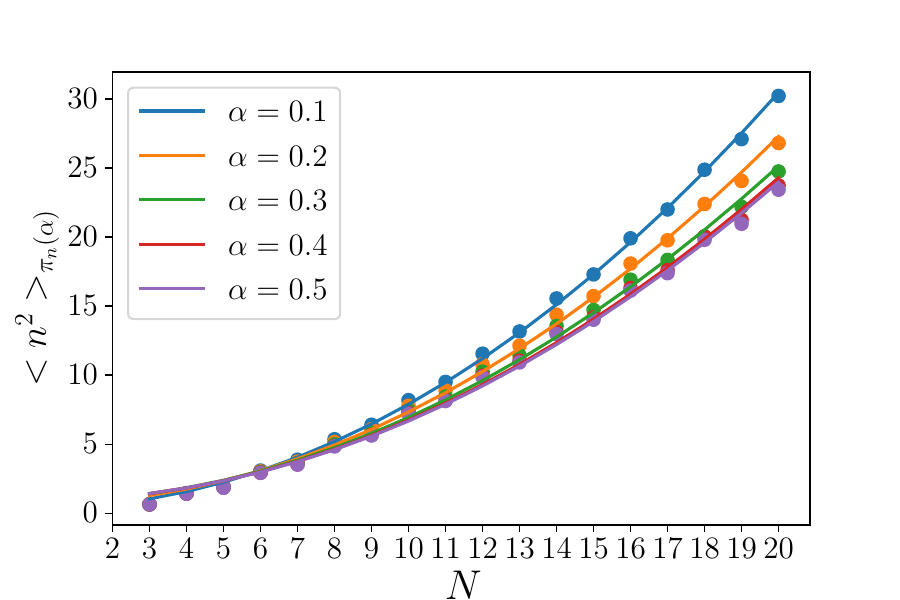}
    \caption{Variance of the walker-position in a quantum walk on a linear quantum network with periodic boundary conditions. The labeled colors show the scaling of the variance with the size of the homogenous network, $N$, for different initial network entanglement values, $\alpha=\{0.1,0.2,0.3,0.4,0.5\}$. The smooth curves are fits of the form, $\braket{n^2}_{\pi_n(\alpha)}=a(\alpha)N^2+b(\alpha)$, with $a(\alpha)=\{0.075,0.066,0.060,0.058,0.057\}$ and $b(\alpha)=\{0.38,0.70,0.89,0.91,0.92\}$ for the 5 different $\alpha$ values.}
    \label{fig:variance}
\end{figure}

The numerical results using both methods for obtaining the stationary distribution of the walker-position probabilities are shown in Fig. \ref{fig:stat_dist_1}. Due to the symmetry of the initial coin state, $\ket{c_0}_c=(\ket{0}+i\ket{1})/\sqrt{2}$, and the initial position being, $n=0$, the expectation value of the position with respect to the stationary distribution is zero - in the stationary distribution. As shown in Fig. \ref{fig:variance}, the variance of the walker-position, $\braket{n^2}_{\pi_n(\alpha)}=\sum_{n}n^{2}\pi_n(\alpha)$, for a QWQN-L on a linear graph with periodic boundary conditions where the nodes are symmetrically labelled, $n=\{-N/2,...,-2,-1,0,+1,+2,...,N/2\}$, can be estimated to be, $\braket{n^2}_{\pi_n(\alpha)}\sim a(\alpha)N^2+b(\alpha)$, where, $a(\alpha),b(\alpha)=\mathcal{O}(1)$, are both functions of $\alpha$ with, $a(\alpha)$ monotonically decreasing with $\alpha$. Together, the first two moments of the walker-position in the stationary distribution for, $\alpha\in[0,0.5]$ and $N\to\infty$, can be summarized as,
\begin{align}
\braket{n}_{\pi_n(\alpha)}&=0,\nonumber\\
\braket{n^2}_{\pi_n(\alpha)}&\sim a(\alpha)N^2.
\label{eq:position_moments}
\end{align}
Since, $a(\alpha)$ decreases with $\alpha$, increasing values of the initial network entanglement localizes the walker about the initial position around the starting position of the walk, as seen in Fig. \ref{fig:stat_dist_1}.

\subsection{Approach to the long-time average probability distribution}
\label{subsec:approach_asymptotic}
\begin{figure}[h]
    \centering
    \includegraphics[height=6cm,width=\columnwidth]{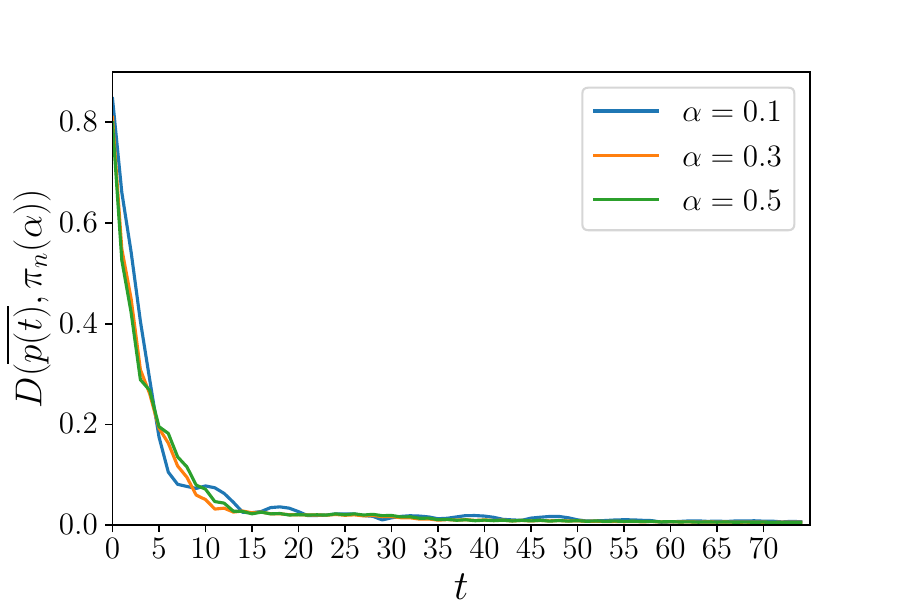}
    \caption{Distance between time-average walker-position probability and the stationary distribution, , vs time. Shown are the results for a QWQN-L on a quantum network with, $N=15$, nodes with periodic boundary conditions, and an unbiased initial coin state, $\ket{c_0}_c=(\ket{0}_{c} + i\ket{1}_{c})/\sqrt{2}$ and starts from the initial position, $n=0$.}
    \label{fig:distance}
\end{figure}

Finally, our numerical investigations reveal that the time, $t_{\pi_n}$, taken to reach the asymptotic distribution under the QWQN-L dynamics is independent of the initial network entanglement, $\alpha$, as shown in Fig. \ref{fig:distance}. We estimate, $t_{\pi}$, as the time after which the distance, $D(\overline{p(t)},\pi_n(\alpha))$, between the time-averaged walker-position probability and the asymptotic distribution with,
\begin{align}
    D(\overline{p(t)},\pi_n(\alpha)) := \frac{1}{2}\sum_{n}|\overline{{p_{n}}} - \pi_{n}(\alpha)|,
\end{align}
is likely to be close to zero. Note that the range, $0\leq D(\overline{p(t)},\pi_n(\alpha))\leq 1$, with the lower bound reflecting complete convergence of the time-averaged probability distribution to the stationary distribution whereas higher values of $D(\overline{p(t)},\pi_n(\alpha))$ denoting correspondingly higher distances.

In Fig. \ref{fig:distance}, the plots of, $D(\overline{p(t)},\pi_n(\alpha))$ vs $t$, for different $\alpha$-values reveal that the approach to the stationary distribution shows higher fluctuations with increasing values of the initial network entanglement, $\alpha$. However, the time $t_{\pi}$ to approach the stationary distribution $\pi_n(\alpha)$ remains independent of $\alpha$ although the stationary distribution itself varies with the latter. The time, $t_{\pi}$, is expected to scale with the size, $N$, of the network since the walker is expected to take an amount of time, $t\sim N/2$, to explore all network vertices in the linear network with periodic boundary conditions. This is due to the unbiased initial coin state resulting in a birectional superposition of walker moves spanning one network-edge per time-step. However, in our simulations with small network sizes up to, $N\leq 15$, we were unable to determine the scaling of $t_{\pi}$ with $N$.

\subsection{Entanglement evolution in the QWQN-L}
\label{subsec:entropy}
The unitary dynamics of the QWQN-L on the initially unentangled degrees of freedom of the quantum network, walker-coin and walker-position, $\ket{\psi(0)}=\ket{\psi}_G\otimes \ket{c_0}_c\otimes \ket{0}_p$, correlates all three degrees of freedom as the walk evolves. The time-dependent state, $\ket{\psi(t)}=\hat{U}_{QWQN}\ket{\psi(0)}=\sum_{i} f_{i}(\alpha) \ket{G_i}\otimes\ket{W_i(t)}$, given by Eq. (\ref{state-qwqn-t}), therefore, is a superposition of multiple product state of the three subsystems. We study the temporal evolution of the entanglement in two bipartite settings.

\begin{figure}[h]
    \centering
    \includegraphics[height=6cm,width=\columnwidth]{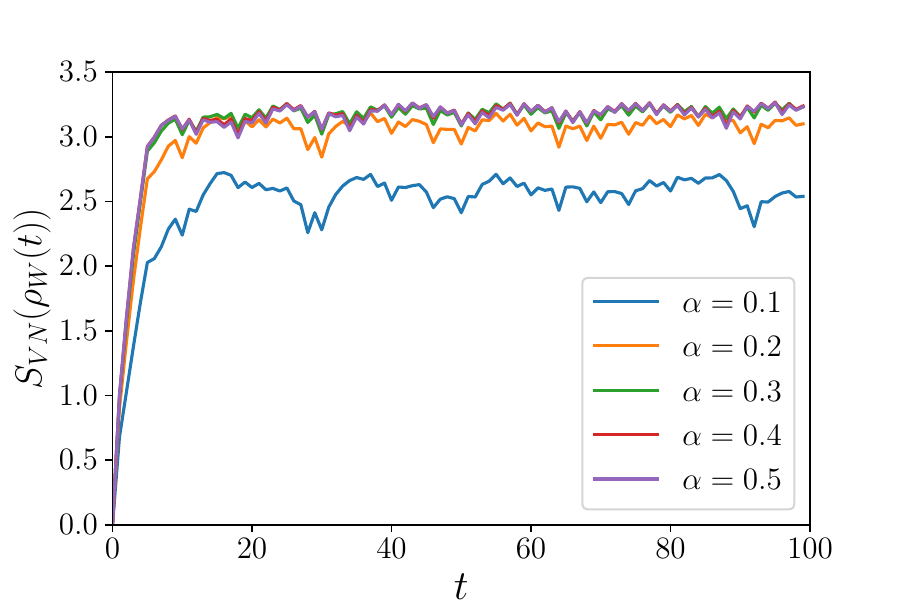}
    \caption{Evolution of the entanglement between the walker degrees of freedom (coin and position) with the quantum network qubits in a network with $N=10$ vertices. The initial coin state is, $\ket{c_0}_c=(\ket{0}_{c} + i\ket{1}_{c})/\sqrt{2}$, and the walker starts from the initial position, $n=0$.}
    \label{fig:Ent_growth_1}
\end{figure}

\emph{Walker-network entanglement}:- First, we look at the bipartite entanglement evolution of system with the quantum network qubits as one subsystem while the walker-coin and walker-position together form the other subsystem. In this case the bipartite entanglement entropy \cite{review_area_laws}, $EE(\ket{\psi(t)})$, of the state $\ket{\psi(t)}$ is given by the von Neumann entropy of the reduced density matrix, $\rho_W(t)$, of the network qubits, that is,
\begin{align}
    EE(\ket{\psi(t)})=S_{VN}(\rho_W(t))=-\tr(\rho_W(t) \log(\rho_W(t))),
\end{align}
with, $\rho_W(t)=\sum_{i=0}^{(2^N-1)}f_i^2(\alpha)\ket{W_i(t)}\bra{W_i(t)}$, given by Eq. (\ref{eq:coin-position_denmat}). A numerical evaluation of this entropy evolution is shown in Fig. \ref{fig:Ent_growth_1} which reveals that $S_{VN}(\rho_W(t))$ approaches a mean saturation value that increases with $\alpha$. On the other hand, the fluctuations about these mean values get suppressed with increasing $\alpha$. Further, this asymptotic behavior of $S_{VN}(\rho_W(t))$ is obtained in a time that approximately matches the time $t_{\pi}$ that the QWQN-L walk dynamics takes to reach the stationary walker-position probability distribution, discussed in relation to Fig. (\ref{fig:distance}).

Interestingly, the form of the state, $\rho_W(t)$, allows us to obtain an upper bound on $S_{VN}(\rho_W(t))$ using the concavity of the entropy function. Since $\rho_W(t)$ is a convex combination of states, $\rho_i(t)=\ket{W_i(t)}\bra{W_i(t)}$, with probability weights, $p_i=f_i^2(\alpha)$, we get that,
\begin{align}
    \sum_i p_i S_{VN}(\rho_i) &\leq S_{VN}(\rho_W(t)) \leq  H(p_i) + \sum_i p_i S_{VN}(\rho_i),\nonumber\\
    \implies0&\leq EE(\ket{\psi(t)})\leq NS_{VN}(\ket{\alpha}),
    \label{eq:ee_lin_bound}
\end{align}
where, $S_{VN}(\rho_i(t))=0$, for any of the conditional walk pure states, $\rho_i(t)$, and, $H(p_i)=-\sum_{_i}f_i^2(\alpha)\log f_i^2(\alpha)$, can be directly calculated from the form of the $f_i(\alpha)$ given in Eq. (\ref{f_i}). Alternately, $H(f_i^2(\alpha))$ can be simply understood as the Von Neumann entropy of $N$-copies of the state, $\alpha$.

The linear scaling of the upper bound on the entanglement entropy $EE(\ket{\psi(t)})$ with the network size, $N$, as shown in Ineq. (\ref{eq:ee_lin_bound}) is rather loose as can be seen from dimensional considerations. To see this, notice that for the state, $\ket{\psi(t)}=\sum_{i} f_{i}(\alpha) \ket{G_i}\otimes\ket{W_i(t)}$, the Hilbert space dimension of the walker-coin and walker-position subsystem is $2N$ whereas that for the quantum network is $2^{N}$. Since the entanglement entropy in a composite system of dimension $d_1\times d_2$ is upper bounded by $\log(\text{min}(d_1,d_2))$ we infer that for $\ket{\psi(t)}$ the maximum entanglement entropy is proportional to $\log(2N)$. From Fig. \ref{fig:Ent_growth_1} we see that the asymptotic mean value of the $EE(\ket{\psi(t)})$ reaches close to this maximum value.

\emph{Quantum network entanglement}:- The QWQN-L dynamics correlates previously uncorrelated quantum network qubits among themselves as described previously in the discussion around Eq. (\ref{state-graphsuper-2}) and (\ref{time-dep-network_state}). The evolution of the correlation between the quantum network qubits can be analysed by studying the bipartite network entanglement of the time-dependent mixed state, $\rho_G(t)=\tr_{c,p}[\ket{\psi(t)}\bra{\psi(t)}]=\sum_{i,j=0}^{(2^N-1)}f_i(\alpha)f_j(\alpha)\braket{W_j(t)|W_i(t)}\ket{G_i}\bra{G_j}$, across a suitable bipartition.

 The bipartite network entanglement, $BNE_{A|B}(\rho_G(t))$, which is the negativity of the mixed-state, $\rho_G(t)$, across a $(A,B)$-bipartition of the network qubits as described in Sec. \ref{subsec:qnets} can be calculated as, 
 \begin{align}
    \mathcal{N}(\rho_G(t)) =\sum_{i=1}^{\text{Dim}(\mathcal{H}_A)} \frac{|\lambda_i(t)| - \lambda_i(t)}{2},
    \label{neg}
\end{align}
where $\lambda_i(t)$'s are the eigenvalues of partial transpose of density matrix $\rho_G(t)$ with respect to the subsystem $A$. The Hilbert-space dimension, $\text{Dim}(\mathcal{H}_A)$, of the subsystem $A$ with $n_A$-qubits is $2^{n_{A}}$. The maximum value that the negativity can take for such a system is $\text{Max}(\mathcal{N}(\rho_G(t)))=(2^{n_A}-1)/2$.

\begin{figure}[h]
    \centering
    \includegraphics[height=6cm,width=\columnwidth]{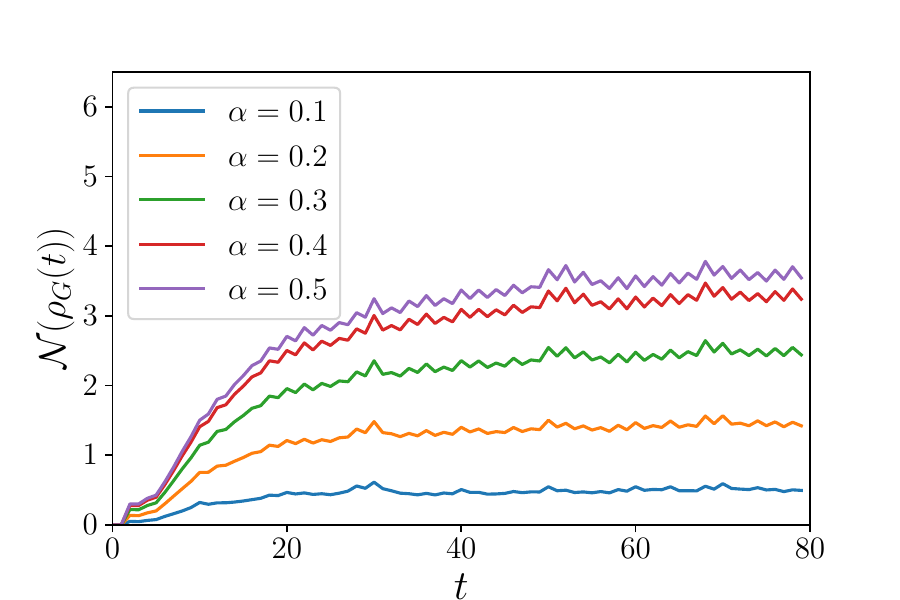}
    \caption{Temporal evolution of the bipartite network entanglement negativity, $BNE(\rho_G(t))$, of the state $\rho_G(t)$ of a quantum network with $N=10$ vertices and an equal bipartition. The initial coin and position state is $\frac{1}{\sqrt{2}}(\ket{0}_{c} + i\ket{1}_{c})\otimes\ket{0}_{p}$. The negativity saturates to mean values increasing with the initial network entanglement $\alpha$.}
    \label{fig:Ent_growth_2}
\end{figure}

The evolution of the bipartite network entanglement, $BNE_{A|B}(\rho_G(t))$ in the QWQN-L in a network with $N=10$ nodes is  shown in Fig. \ref{fig:Ent_growth_2} for an equipartition of the quantum network, $(A,B)$ with 10 qubits each. The initial negativity is zero since the biparition was chosen to be across the $l_1$-type boundary as shown in Fig. \ref{fig:QN}. For such a bipartition, the boundary $l_1$ separates qubits into two groups with no entangled pair straddling the boundary at time $t=0$. The value of negativity at large times, $t\to\infty$, can be seen to saturate around a mean value that increases with the initial network entanglement, $\alpha$, and with correspondingly higher fluctuations. Notice that the maximum bipartite network entanglement of the network qubits, $\text{Max}(\mathcal{N}(\rho_G(t)))\approx 3$, corresponding to the largest value of $\alpha=0.5$ is still significantly smaller than the maximum value of negativity, $(2^{10}-1)/2=511$, obtainable for this network with $20$ qubits across an equipartition with $n_A=n_B=10$. The reason for this rather small value of the maximum bipartite network entanglement is that the correlation between the $A$ and $B$ parts of the network is mediated via the walker degrees of freedom which have a much smaller Hilbert-space dimension than that of $A$ or $B$. The walker-coin and walker-position degrees of freedom have a dimension of $2N$ - that is, linear in the network size whereas the dimension of $A,B$ is $2^{N}$.


\subsection{Interference among the quantum walk state amplitudes}
\label{subsec:interference}
We now verify that the model of QWQN-L described above indeed exhibits the central feature of a quantum walk, which is, interference among the amplitudes of the quantum walk state, $\ket{\psi(t)}$. This is a necessary criterion for the walk dynamics in the QWQN-L to be \emph{quantum} \cite{kendon_qw_criteria}. To see that interference does happen in the QWQN-L it is useful to keep in mind the form of the time-dependent state of the QWQN-L, $\ket{\psi(t)}=\sum_{i} f_{i}(\alpha) \ket{G_i}\otimes\ket{W_i(t)}$, as described in Eq. (\ref{physical_pic_state}). And in such a state, consider a conditional quantum walk state, $\ket{W_i(t)}$, corresponding to the computational basis state $\ket{G_i}$ and iterate the associated walk operator, $\hat{U}^i$, defined via Eq. (\ref{eq:u_i}), for a few number of time-steps.

Interference occurs between the amplitudes, $a^i_n(t),b^i_n(t)$, of the different coin and position components of a given conditional walk-state, $\ket{W_i(t)}=\sum_n (a^i_n(t)\ket{0}_c+b^i_n(t)\ket{1}_c)\otimes \ket{n}_p$, defined in Eq. (\ref{walk_states}). This is most easily seen when the quantum network basis state $\ket{G_i}$ corresponding to $\ket{W_i(t)}$ happens to be $\ket{G_i}=\ket{01}^{\otimes N}$ in which the parity of the two qubits at every vertex in the network is odd. Thus, the conditional unitary applied to the walker-coin at each vertex is the Hadamard unitary operator as discussed in relation to Eq. (\ref{eq:parity_operators}). The conditional walk $\ket{W_i(t)}$ in this case, therefore, is identical to a standard quantum random walk, DCQW-L. 
Following this specific walk state, $\ket{W_i(t)}$, for the first three time-steps we obtain,
\begin{align}
\ket{W_i(0)}&=(\frac{1}{\sqrt{2}} \ket{0}_c +\frac{i}{\sqrt{2}} \ket{1}_c) \otimes \ket{0}_p,\\
\ket{W_i(1)}&=\frac{(1+i)}{2}\ket{0}_c \otimes \ket{1}_p + \frac{(1-i)}{2}\ket{1}_c \otimes \ket{-1}_p, \\
\ket{W_i(2)}&=\frac{(1+i)}{2\sqrt{2}}\ket{0}_c \otimes \ket{2}_p \nonumber\\
+&\frac{(1+i)}{2\sqrt{2}}\ket{1}_c \otimes \ket{0}_p+\frac{(1-i)}{2\sqrt{2}}\ket{0}_c \otimes \ket{0}_p\nonumber\\
-&\frac{(1-i)}{2\sqrt{2}}\ket{1}_c \otimes \ket{-2}_p.
\label{onestepint}
\end{align}
The second equation above shows the interference between the amplitudes of the $\ket{0}_c\ket{0}_p$ and $\ket{1}_c\ket{0}_p$ states. Similarly, the third equation shows the interference between the amplitudes of the $\ket{0}_c\ket{1}_p$ and the $\ket{1}_c\ket{-1}_p$ states.

There are also conditional walk states, $\ket{W_j(t)}$, contributing to the time-dependent state, $\ket{\psi(t)}=\sum_{i} f_{i}(\alpha) \ket{G_i}\otimes\ket{W_i(t)}$, which exhibit no quantum walk-like behavior. This can be seen clearly by considering the walk state, $\ket{W_j(t)}$, corresponding to the quantum network state $\ket{G_j}=\ket{00}^{\otimes N}$, which has even-parity of the two qubits at every vertex of the network. Thus, the conditional unitary applied to the walker-coin at each vertex is the identity operator as discussed in relation to Eq. (\ref{eq:parity_operators}). Therefore, in the state, $\ket{W_j(t)}$, the walker moves ballistically on the linear quantum network without any splitting of its amplitude.

The overall quantum walk in the QWQN-L is a coherent superposition of independent quantum walks, $\ket{W_i(t)}$, conditioned on the quantum network states, $\ket{G_i(t)}$, weighted by the functions, $f_i(\alpha)$, which pertains to the initial network entanglement. Each independent quantum walk can be understood to have different degrees of \emph{quantum walk-like} behavior based on the associated quantum network state. Overall, we conclude that the QWQN-L fulfils the criteria of interference between its component amplitudes for the QWQN-L to be considered a genuine quantum walk.

\subsection{Quantum walk statistics as probes of network entanglement}
\label{subsec:network_probe}

We have seen so far that the stationary distribution, $\pi_n(\alpha)$, of the walker-position described in Subsec. \ref{subsec:stationary_dist} depends on the initial network entanglement, $\alpha$, in the case of homogenous networks where all edges have the same value of their entanglement measure. This raises the intriguing possibility whether for inhomogenous networks the stationary distribution can also reveal some aspects of the network entanglement.

\begin{figure}[t]
\subfloat[Comparison of stationary probability distribution of the walker position for heterogenous and homogenous networks. (Square data-marker and solid connecting lines ) Stationary distribution, $\pi_n(\overline{\alpha})$, in inhomogenous network, for different $\overline{\alpha}$ values and fixed standard-deviation $\sigma_\alpha=20\%$ of the maximum allowed. (Circle data-markers and dashed connecting lines) Stationary distribution, $\pi_n(\overline{\alpha})$, in a homogenous network with all $\alpha=\overline{\alpha}$.]{%
  \includegraphics[clip,width=\columnwidth]{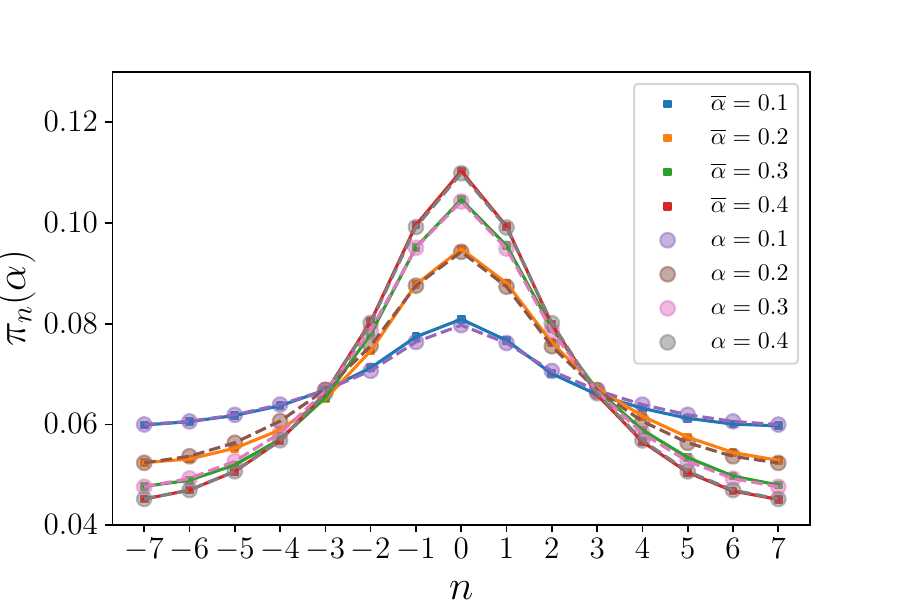}%
}

\subfloat[Stationary distribution, $\pi_n(\overline{\alpha})$, in inhomogenous network, for different $\overline{\alpha}$ values and different standard-deviations $\sigma_\alpha=20\%,50\%,80\%$ of the maximum allowed for that $\alpha$.]{%
  \includegraphics[clip,width=\columnwidth]{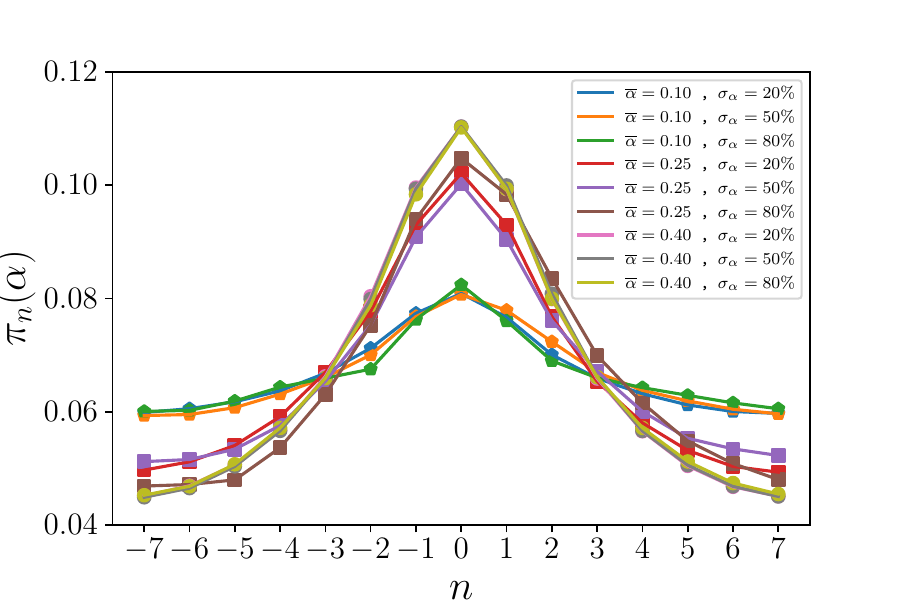}%
}

\caption{Stationary probability distribution, $\pi_n(\overline{\alpha})$, for the walker-position, $n$, in the QWQN-L for different values of the average initial network entanglement, $\overline{\alpha}$, in inhomogenous networks. The quantum network has $N=15$ nodes with periodic boundary conditions, and an unbiased initial coin state, $\ket{c_0}_c=(\ket{0}_{c} + i\ket{1}_{c})/\sqrt{2}$ and starts from the initial position, $n=0$. The top and bottom panels respectively show $\pi_n(\overline{\alpha})$ for relatively small and large values of the standard-deviation. Comparison of the two panels shows qualitatively the robustness of the distribution to significant deviations in the inhomogeneity of edge entanglement values.}
\label{fig:inhom_network}
\end{figure}

Our numerical investigations with inhomogenous quantum networks, shown in Fig. \ref{fig:inhom_network}, reveal that the stationary distribution in this case depends on the average of the entanglement over the network edges, $\overline{\alpha}=(\sum_{i=1}^N\alpha_i)/N$. Each inhomogenous network utilised in the simulations was chosen from a realization with a fixed average and standard deviation of the edge entanglement values. The average entanglement values were chosen to be $\overline{\alpha}\in\{0.1,0.2,0.3,0.4\}$ and for any given $\overline{\alpha}$ the standard deviations for network realizations, $\sigma_\alpha$, was taken to be $20\%,50\%$ and $80\%$ of the maximum allowed for that value of $\overline{\alpha}$. 

From the top panel in Fig. \ref{fig:inhom_network} we see that the stationary distribution in an inhomogenous network with $N=15$ vertices reveals that a higher value of $\overline{\alpha}$ leads to a localization of the stationary distribution, $\pi_n(\overline{\alpha})$, around the initial position of the walker similar to the homogenous case shown in Fig. \ref{fig:stat_dist_1}. Importantly, we see that the stationary distribution for the inhomogenous and homogenous networks have a high degree of similarity, that is, $\pi_n(\alpha)=\pi_n(\overline{\alpha})$ for $\overline{\alpha}=\alpha$, in networks with modest amount of inhomogeneity, $\sigma_{\alpha}\sim 20\%$, in the edge entanglement values. 

The bottom panel in Fig. \ref{fig:inhom_network} shows that the qualitative behavior of the stationary distribution stays the same even in inhomogenous networks with significantly high standard deviations in their edge entanglement values. From the same panel, we further note that the role of the standard deviation of the edge entanglement values in determining the form of $\pi_n(\overline{\alpha})$ is suppressed whenever the average entanglement is close to its allowed limits, $\overline{\alpha}\to0$, or, $\overline{\alpha}\to 0.5$. Whereas, for $\overline{\alpha}\approx0.25$, which is the mid-point of its range of allowed values, large standard deviation of the edge entanglement values gets reflected in the variance of the stationary distributions for the different inhomogenous networks.

These results for inhomogenous networks suggest a quantum walk-based method for inferring the average network entanglement, $\overline{\alpha}$. This can be achieved by measuring the stationary occupation probability of the walker at the initial position, $\pi_{n=0}^{meas}$, and comparing it with the numerically simulated value, $\pi_{n=0}(\overline{\alpha})$ for a range of standard-deviation $\sigma_\alpha$ values, for a given network size, $N$. Since, $\pi_{n=0}^{meas}=\lim_{T\to\infty}(1/T)\sum_{t}p_{n=0}(t)$, one needs to measure the occupation probabilities in the $n=0$ position at $T$-different instants of time, $t=t_1,t_2,...,t_T$. Assuming that $m_w$-measurements are done for each value of $t$, the total number of measurements required is $m_wT$. A reasonable choice of the averaging time, $T$, is the time required for the QWQN-L dynamics to approach the asymptotic stationary distribution which was shown to scale as, $T\sim\mathcal{O}(N)$, in Subsec. \ref{subsec:approach_asymptotic}. Therefore, with, $N^{meas}_{QW}=\mathcal{O}(m_wN)$, measurements one can estimate the average initial entanglement in the network, $\overline{\alpha}$, with a standard error that scales as $\sim (1/m_w)$. 

The number of measurements in the quantum walk-based method is comparable to that in a direct estimate of the average entanglement, $\overline{\alpha}=(\sum_e \alpha_e)/N$, that requires measuring the entanglement over each of the $N$-edges in the network. The total number of measurements in the direct method is, $N^{meas}_{dm}=m_eN$, where $m_e$ is the number of measurements required to estimate the entanglement of the state over a single edge of the network. However, a closer inspection of the two methods reveals an advantage in the number of required measurements in the quantum walk-based method.
This is because the direct method requires the quantum state tomography of a two-qubit system needing about, $m_e\sim (10^5-10^6)$, measurements \cite{two_qubit_tomo} over each edge and even then the fidelity of the reconstructed state is about $0.75$. This makes the direct method significantly more resource-intensive, $m_e\gg m_w$, than the quantum walk-based method which only relies on relatively simple projective measurements of the walker at the position, $n=0$. A caveat here is that the quantum walk-based method is likely to work better for networks with lower heterogeneity in the edge entanglement values as depicted in the two panel of Fig. \ref{fig:inhom_network}.

\section{Discussion and conclusions}
\label{sec:disc_conc}
In this work we have described a novel type of a quantum random walk, the QWQN-L, in which the underlying quantum network plays an active role in the walk evolution - distinct from the passive role of the underlying classical graph in canonical quantum walks \cite{andraca_review, kadian_review}. In the latter, the graph essentially only provides a set of vertices for the walker moves, whereas, in the QWQN-L the quantum network qubits interact with the walker degrees of freedom via conditional unitary operations based on the state of the network qubits at every vertex. These conditional unitaries correlate the quantum network qubits with the walker degrees of freedom, viz., the walker-coin and walker-position. The overall unitary dynamics in the QWQN-L results in a time-dependent pure state of the whole system that can be interpreted as a coherent superposition of multiple independent conditional quantum walks - each conditioned on a different computational basis state of the quantum network.

We have shown that the randomness of the walker-coin state in the QWQN-L, essential for the randomness of the walker moves, emerges from the entanglement of the quantum network qubits among themselves. In the limit of a completely unentangled quantum network there is no randomness of the walker behavior in the QWQN-L and the walker moves ballistically over the network vertices. Whereas, in the limit of pair-wise maximally-entangled network qubits, the QWQN-L is a coherent  superposition of quantum walks each with varying degrees of ballistic and diffusive behavior depending on the associated computational basis state of the network qubits. Since the overall dynamics of the QWQN-L is unitary, the time-dependent walker probability distribution over the network vertices does not converge to a stationary distribution. However, the asymptotic time-averaged probability distribution does converge to a stationary distribution similar to the scenario of quantum equilibration under unitary dynamics \cite{loschmidt_echo}. We saw that the stationary walker-position probability distribution shows increasing localization around its initial position with higher values of the initial network entanglement. In the stationary distribution, the average displacement of the walker-position is zero, $\braket{n}_\alpha=0$, whereas, the variance scales quadratically, $\braket{n^2}_{\alpha\to0}\sim N^2$, for, $\forall \alpha\in[0,0.5]$.

Further, we studied the growth of the bipartite entanglement between the network qubits across a natural bipartition as well as the entanglement between the walker and the network qubits due to the QWQN-L dynamics. The asymptotic values for both these entanglement measures increase with the network entanglement at the initial time. The time to approach the asymptotic values for all these measures was found to be effectively independent of the initial network entanglement; and we conjecture depends only on the network size. Finally, we noted the interference between the amplitudes of the network-coin-position states contributing to the time-dependent QWQN-L state showing indeed that the QWQN-L satisfies the quantum walk criteria \cite{kendon_qw_criteria}.

While the QWQN-L is interesting in itself as a novel model of a quantum walk where the randomness of walker moves is sourced by the entanglement of the underlying network, we also pointed out a potential application in estimating the average entanglement in inhomogenous quantum networks. Such an application utilises the fact that the time-averaged walker-position probability distribution reflects the average amount of the initial network entanglement and can be used to infer it from the measured time-averaged occupation probability of the walker at the initial position. We saw that this can significantly reduce the complexity of estimating the average entanglement in a network with a simple ring topology.

This work presents a new direction in the use of quantum networks: as underlying graphs that can be potentially used to achieve target quantum walks statistics for various information processing tasks \cite{andraca_review,kadian_review}. 
Interesting extensions of this work fall primarily into two categories - first, that involve generalising the properties of the underlying quantum network; second, those that generalise the properties of the quantum walker.
In the first category, more general network topologies including regular planar lattices and random networks \cite{qn_topography} and long-range initial network entangled states can be considered. In the second category, more general coin and shift operators \cite{andraca_review} that can extract more information about the entanglement properties of the quantum network would be desirable. In future work, we will develop these aspects of quantum walks on quantum networks altogether as an accurate tool to probe refined entanglement properties of general quantum networks.

\emph{Acknowledgement}:- Funding from DST, Govt. of India through the SERB grant MTR/2022/000389, IITB TRUST Labs grant DO/2023-SBST002-007 and the IITB seed funding is gratefully acknowledged. 

\appendix

\section{Fourier transform of the QWQN-L state}
\label{app:ft}
Here, we consider the QWQN-L walk dynamics in the Fourier transformed space as is done in the treatment of the canonical quantum walk, DCQW-L. In the latter the translational invariance of the walk unitary over the graph vertices leads to conserved momentum making the unitary diagonal in the momentum basis which is not the case in the QWQN-L as mentioned in the main text.  Our goal is to highlight the step at which the analytical situation of the QWQN-L deviates from that in the DCQW-L. 

Towards this goal, we consider the Fourier transform of the position states, $\ket{j}, j\in V$, of the walker given by, 
\begin{align}
\ket{\Tilde{k}} = \frac{1}{\sqrt{N}}\sum_{j=0}^{N-1}\omega_{N}^{j\tilde{k}}\ket{j}
\label{FT_n}
\end{align}
where, $\omega_{N} = \exp{\frac{2\pi i}{N}}$, and, $\ket{\tilde{k}}, \tilde{k}=0,...,(N-1)$, are the new basis states which we call the momentum states. In the following we obtain the action of the walk unitary, $\hat{U}_{QWQN}$, given by Eq. (\ref{eq:u-qwqn}), on states of the form, $\ket{G_i}\ket{c}\ket{\tilde{k}}$, to understand the walk dynamics in this graph-coin-momentum basis. 

Notice that the time-dependent walk-state, $\ket{\psi(t)}$, given by Eq. (\ref{state-qwqn-t}) may be rewritten as,
\begin{align}
\ket{\psi(t)} = \sum_{j=0}^{N-1}\sum_{c=0}^{1}\sum_{i=0}^{2^{N}-1}f_{i}(\alpha)\psi_{i,c,j}(t)\ket{G_{i}}\ket{c}\ket{j},
\label{psit}
\end{align}
with, $\psi_{i,0,j}(t)=a^i_n(t),\psi_{i,1,j}(t)=b^i_n(t)$, and applying the Fourier transform in Eq. (\ref{FT_n}) to the position states, $\ket{j}$, above we get,
\begin{align}
\ket{\Tilde{\psi}(t)} = \sum_{\tilde{k}=0}^{N-1}\sum_{c=0}^{1}\sum_{i=0}^{2^{N}-1}f_{i}(\alpha)\Tilde{\psi}_{i,c,k}(t)\ket{G_{i}}\ket{c}\ket{\tilde{k}}.
\label{psitilde}
\end{align}

The single time-step action of the walk unitary in the $\ket{G_i}\ket{c}\ket{\tilde{k}}$-basis is given by,
\begin{align}
&\ket{\tilde{\psi}(t+1)}=\hat{U}_{QWQN}\ket{\tilde{\psi}(t)},\nonumber\\
&=(\openone_G\otimes \hat{S}_{c\to p} )(\hat{H}_{G\to c}\otimes \openone_p)\ket{\tilde{\psi}(t)},\nonumber\\
&=\sum_{\tilde{k},c,i} f_{i}(\alpha)\Tilde{\psi}_{i,c,k}(t)(\openone_G\otimes \hat{S}_{c\to p} )(\hat{H}_{G\to c}\otimes \openone_p)\ket{G_{i}}\ket{c}\ket{\tilde{k}},
\end{align}
which requires us to consider the action of $\hat{U}_{QWQN}$ on a basis state, $\ket{G_i}\ket{c}\ket{\tilde{k}}$ as follows,
\begin{align}
&\hat{U}_{QWQN}\ket{G_i}\ket{c}\ket{\tilde{k}}= \hat{S}(\hat{H}_{G\to c}) \ket{G_i}\ket{c}\ket{\tilde{k}}\nonumber\\
&=\hat{S}_{c\to p}( (\sum_{n\in V}(P^e_{n_ln_r}\otimes \openone_c+P^o_{n_ln_r}\otimes \hat{H}_c) \ket{G_i}\ket{c}\ket{\tilde{k}}).
\label{Utopsitilde}
\end{align}
Note in the second line above, the parity operators $P^{o}_{n_l,n_r},P^{e}_{n_l,n_r}$ conditionally apply either the identity or the Hadamard unitary operator on the coin $\ket{c}$ based on the parity of the qubits at position $n$ in the given computational basis state, $\ket{G_i}$, of the network. Since, $\ket{\tilde{k}}$ is a sum over the position states it is clear that $\ket{G_i}\ket{c}\ket{\tilde{k}}$ is not an eigenstate of $\hat{H}_{G\to c}$ and hence of the operator $\hat{U}_{QWQN}$. Therefore, the graph-coin-momentum basis does not diagonalise the walk-unitary and techniques from DCQW-L are no longer relevant. 

Intuitively, one can summarize the above discussion to state that because the coin operator at every vertex is conditioned on the state of the network qubits at that vertex there is no translational invariance of the walker dynamics. This results in no conserved momentum that could have enabled a solution in the Fourier transform basis.

\bibliographystyle{apsrev4-1}
\bibliography{refs-qnet.bib}  
\end{document}